\documentclass[aps,prm,reprint,amsmath,amssymb,superscriptaddress,longbibliography, bibnotes]{revtex4-2}

\usepackage{graphicx}
\usepackage{xcolor}
\usepackage{bm}
\usepackage{hyperref}
\usepackage{amsmath}
\usepackage{amssymb}
\usepackage{physics}
\usepackage{siunitx}
\usepackage{booktabs}
\usepackage{multirow}
\usepackage{dcolumn}
\usepackage{url}
\usepackage{makecell}
\usepackage{tikz}
\usetikzlibrary{arrows.meta, backgrounds, calc, fit, positioning}

\newcommand{\vR}{\mathbf{R}}
\newcommand{\vk}{\mathbf{k}}

\newcommand{\Mz}{\mathcal{M}_{z}}

\newcommand{\kk}{\mathbf{k}}

\newcommand{\paoflow}{\textsc{PAOFLOW}}
\newcommand{\code}[1]{\texttt{#1}}

\newcommand{\anticomm}[2]{\bigl\{ #1 , #2 \bigr\}}

\begin{document}

\title{\paoflow{}: an automated suite for
       \textit{ab initio} electronic, transport, and topological
       properties of materials}

\author{Anooja Jayaraj}
\affiliation{Empa, Swiss Federal Laboratories for Materials Science and Technology, nanotech@surfaces Laboratory, Überlandstrasse 129, CH-8600 Dübendorf, Switzerland}

\author{Sergio Alvarruiz}
\affiliation{Zernike Institute for Advanced Materials, University of Groningen, Nijenborgh 3, 9747 AG Groningen, The Netherlands}

\author{Mia Falatko}
\affiliation{Zernike Institute for Advanced Materials, University of Groningen, Nijenborgh 3, 9747 AG Groningen, The Netherlands}

\author{Zhiren He}
\affiliation{Zernike Institute for Advanced Materials, University of Groningen, Nijenborgh 3, 9747 AG Groningen, The Netherlands}
\affiliation{Department of Physics, University of North Texas, Denton, TX 76203, USA}

\author{Jonathan Red}
\affiliation{Department of Physics, University of North Texas, Denton, TX 76203, USA}

\author{Caua F. Schuch}
\affiliation{Escola de Engenharia de Lorena - DEMAR, Universidade de São Paulo, 12612-550, Lorena, Brazil}

\author{Karma Tenzin}
\affiliation{Zernike Institute for Advanced Materials, University of Groningen, Nijenborgh 3, 9747 AG Groningen, The Netherlands}
\affiliation{Department of Physical Science, Sherubtse College, Royal University of Bhutan, Kanglung, Trashigang, Bhutan}

\author{Chao Chen Ye}
\affiliation{Zernike Institute for Advanced Materials, University of Groningen, Nijenborgh 3, 9747 AG Groningen, The Netherlands}

\author{Davide Ceresoli}
\affiliation{Consiglio Nazionale delle Ricerche (CNR) — Istituto di Scienze e Tecnologie Chimiche ``Giulio Natta'' (SCITEC), via Golgi 19, 20133 Milano, Italy}
             
\author{Marcio Costa}
\affiliation{Department of Physics, University of North Texas, Denton, TX 76203, USA}
\affiliation{Instituto de Física, Universidade Federal Fluminense, Niterói, Rio de Janeiro, Brazil}

\author{Stefano Curtarolo}
\affiliation{Department of Mechanical Engineering and Materials Science, Duke University, Durham, NC 27708, USA}

\author{Jagoda S{\l}awi{\'n}ska}
\affiliation{Zernike Institute for Advanced Materials, University of Groningen, Nijenborgh 3, 9747 AG Groningen, The Netherlands}

\author{Marco Buongiorno Nardelli}
\email{mbn@unt.edu}
\affiliation{Department of Physics, University of North Texas, Denton, TX 76203, USA}
\affiliation{Santa Fe Institute, Santa Fe, NM 87501, USA}

\date{\today}

\begin{abstract}
High-throughput first-principles property calculations are often constrained by costly post-processing and dense Brillouin-zone sampling, impeding the creation of large, internally consistent materials-property datasets and limiting AI-driven discovery workflows. Pseudo-atomic-orbital (PAO) Hamiltonians provide an exact tight-binding representation of first-principles electronic structure that enables the calculation of a wide range of electronic, optical, topological, and transport properties at negligible cost, thereby supporting scalable generation of training-quality data and AI-ready data infrastructures. In this work, we present \paoflow{~3.0} -- an open-source Python suite that automates the construction and analysis of PAO Hamiltonians from plane-wave density functional theory calculations performed with either Quantum ESPRESSO or VASP. The resulting Hamiltonians enable efficient electronic structure interpolation, Fermi surface analysis, optical and dielectric response, transport coefficients, Berry phase and topological quantities, quantum transport, and other materials properties. Compared with previous releases, \paoflow{~3.0} substantially extends the scope of the package through the introduction of internal projections enabling support for VASP calculations, self-consistent Hubbard U and V corrections obtained using ACBN0 and eACBN0 methods, generation of environment-dependent Slater-Koster tight-binding models, Landauer--Büttiker quantum transport, and calculation of quantum oscillations using the integrated \textsc{PySKEAF} module. The theoretical foundations and the software architecture are presented together with representative calculations illustrating the current capabilities of the package.
\end{abstract}

\maketitle

\tableofcontents

\section{Introduction}
\label{sec:introduction}

The accurate first-principles prediction of materials properties has become an indispensable tool of modern condensed matter physics and materials science. While conventional plane-wave density functional theory (DFT) calculations provide access to the electronic structure of materials~\cite{Hohenberg1964, Kohn1965}, the evaluation of many experimentally relevant observables often requires computationally demanding post-processing and extensive Brillouin zone (BZ) sampling~\cite{Yates2007}. 

This limits the creation of large, internally consistent materials property datasets and constrains high-throughput workflows designed to leverage modern artificial intelligence (AI) for accelerated materials discovery and rational design~\cite{2019Alberi,Roche2026}. By contrast, pseudo-atomic-orbital (PAO) Hamiltonians, constructed by projecting first-principles wavefunctions onto a reduced basis of localized atomic orbitals, provide an efficient route for evaluating a broad range of materials properties at a negligible computational cost while preserving the accuracy of the first-principles electronic structure~\cite{Agapito2016_1, Agapito2016_2, DAmico2016}. 

\paoflow{} implements this methodology through an automated construction of PAO Hamiltonians directly from first-principles calculations~\cite{paoflow1, paoflow2}. Since its initial release in 2016, the code has evolved considerably from a post-processing tool for plane-wave DFT calculations into a comprehensive Python library for electronic structure analysis and materials properties calculations. The current version of \paoflow{~(3.0)} substantially broadens the scientific scope of the package by enabling the study of emerging physical phenomena and expanding the range of experimentally relevant materials properties accessible from first-principles electronic structure. The capabilities of \paoflow{} span several areas of first-principles materials modeling, including:

\begin{itemize}
  \item \textit{Band interpolation} along arbitrary
    $\mathbf{k}$-paths or throughout the full BZ, density of states (total and projected), Fermi surfaces (\autoref{sec:paoflow});
  \item \textit{Optical and dielectric response}: complex dielectric tensor,
    optical conductivity, joint density of states, electron energy-loss
    spectra (EELS), and complex refractive index (\autoref{sec:paoflow});
  \item \textit{Self-consistent Hubbard corrections}: the ACBN0 and eACBN0
    pseudohybrid density functionals for calculations of U and V parameters in the DFT$+U$ and DFT$+U+V$ methods
    (\autoref{sec:acbn0:acbn0});
  \item \textit{Environment-dependent tight-binding models}:
    Slater-Koster parameterization with structural transferability
    (\autoref{sec:SKTB});
  \item \textit{Semiclassical transport}: electrical and thermal conductivity, Hall and Nernst effects, Rashba-Edelstein effect, as well as user-extensible relaxation time models (\autoref{sec:transport});
  \item \textit{Kubo-formula response}: anomalous Hall, spin Hall, orbital
    Hall conductivities based on Berry curvature integration, and their decomposition as layer-projected properties (\autoref{sec:transport});
  \item \textit{Landauer-Büttiker quantum transport}: transmission functions,
    conductance, and current-voltage characteristics for nanoscale
    conductor/lead geometries (\autoref{sec:transport});
  \item \textit{Lattice dynamics and thermal properties}: phonon dispersions, IR and Raman spectra and thermal properties via an interface with the \textsc{phonopy} package
  (\autoref{sec:phonons});
  \item \textit{Electron-phonon coupling from pseudo-atomic-orbitals interpolation}: electron-phonon coupling, Eliashberg function, superconducting transition temperature
  (\autoref{sec:elphon});
  \item \textit{Quantum oscillation analysis}: de~Haas-van~Alphen (Shubnikov - de~Haas) frequencies and effective masses via the embedded \textsc{pyskeaf} Fermi surface extreme orbit finder (\autoref{sec:topological});
   \item \textit{Electronic structure and topological characterization}: Chern numbers,
    $\mathbb{Z}_2$ indices, Weyl points search and chirality, and band unfolding (\autoref{sec:topological}).
\end{itemize}

In parallel, several software packages have been developed for the calculation of materials properties based on the first-principles electronic structure. Wannier90 is widely used for generation of  maximally-localized Wannier functions (MLWFs)~\cite{Marzari2012,Pizzi2020}. WannierTools and WannierBerri extend its capabilities to topological properties, surface states, and Berry-curvature-related quantities~\cite{Wu2018,Tsirkin2021,Zhi2022}. BoltzTraP2 provides semiclassical transport calculations~\cite{Madsen2018}, while Z2Pack allows the computation of topological invariants~\cite{Gresch2017}. The SKEAF code~\cite{Rourke2012} is a widely used tool for extracting
de~Haas-van~Alphen oscillation frequencies from calculated Fermi surfaces. \paoflow{} brings together most of these capabilities within a single open-source package for first principles materials modeling and adopts a Hamiltonian construction methodology based on deterministic projection onto pseudo-atomic orbitals. The construction of PAO Hamiltonians requires no additional information beyond the wavefunctions obtained from the first principles calculation, making the methodology particularly well-suited for automated and high-throughput materials calculations. In addition, \paoflow{} supports both Quantum ESPRESSO~\cite{Giannozzi2009,Giannozzi2017} and VASP~\cite{Kresse1996,Kresse1999}, two widely used plane-wave DFT codes. This allows the same PAO-based methodology to be applied across different first-principles workflows while preserving a consistent approach to the subsequent calculation of materials properties.



\paoflow{} is freely available at
\url{https://github.com/marcobn/PAOFLOW} under the GNU General Public
License v3 and can be installed from the PyPi repository (\texttt{pip install \paoflow{}}). Documentation, examples and tutorials can be found at \url{https://paoflow.org}.

In this paper, we give a comprehensive description of the
theoretical foundations and software architecture
(\autoref{sec:paoflow}), the ACBN0 and eACBN0 pseudohybrid functionals
(\autoref{sec:acbn0:acbn0}), the environment-dependent tight-binding module
(\autoref{sec:SKTB}), transport properties (\autoref{sec:transport}), phonon calculations (\autoref{sec:phonons}), electron-phonon coupling (\autoref{sec:elphon}) and
the electronic structure and topological properties
(\autoref{sec:topological}). Computational details are listed in \autoref{sec:computation} and technical aspects of software architecture are detailed in \autoref{sec:architecture}. The conclusions are presented in \autoref{sec:conclusions}.

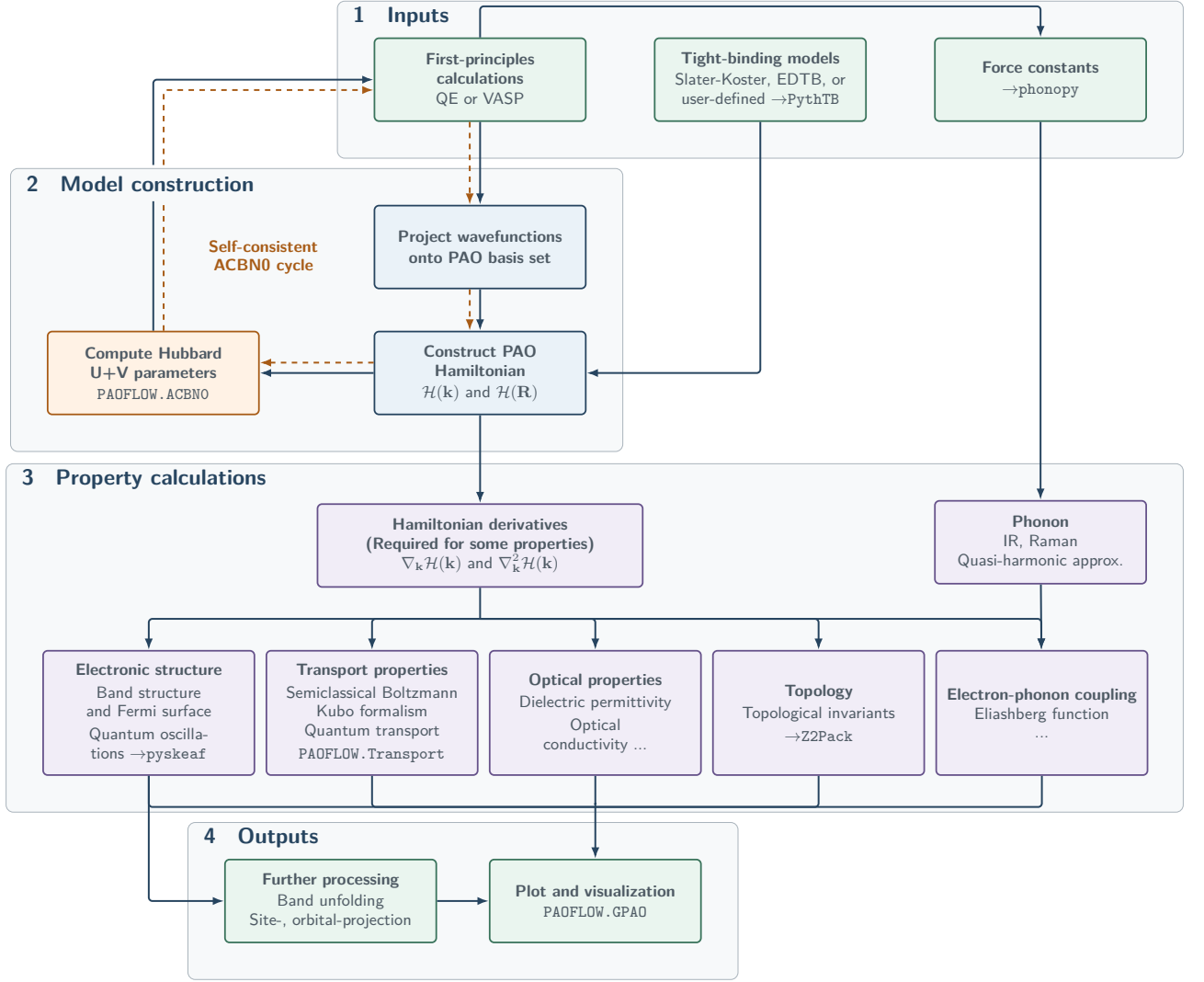
\begin{figure*}[t]
    \centering
    \resizebox{\linewidth}{!}{\definecolor{navy}{HTML}{26445F}
\definecolor{bluefill}{HTML}{EAF2F8}
\definecolor{green}{HTML}{39735B}
\definecolor{greenfill}{HTML}{EAF5EF}
\definecolor{orange}{HTML}{A75B16}
\definecolor{orangefill}{HTML}{FFF2E5}
\definecolor{purple}{HTML}{66528A}
\definecolor{purplefill}{HTML}{F1EDF7}
\definecolor{graytext}{HTML}{4D5963}
\definecolor{grayline}{HTML}{AAB4BC}
\definecolor{stagefill}{HTML}{F7F9FA}

\begin{tikzpicture}[
    font=\sffamily,
    node distance=8mm and 12mm,
    >={Latex[length=2.2mm,width=1.6mm]},
    flow/.style={
        -{Latex[length=2.2mm,width=1.6mm]},
        line width=1pt,
        draw=navy,
        rounded corners=2pt
    },
    flow_noarrow/.style={
    	line width=1pt,
    	draw=navy,
    	rounded corners=2pt
    },
    cycle/.style={
    	-{Latex[length=2.2mm,width=1.6mm]},
    	dashed,
    	line width=1pt,
    	draw=orange,
    	rounded corners=2pt
    },
    cycle_noarrow/.style={
    	dashed,
    	line width=1pt,
    	draw=orange,
    	rounded corners=2pt
    },
    process/.style={
        rectangle,
        rounded corners=2.5pt,
        draw=navy,
        line width=0.9pt,
        fill=bluefill,
        text=graytext,
        align=center,
        minimum height=16mm,
        text width=38mm,
        inner sep=4pt
    },
    input/.style={
        process,
        draw=green,
        fill=greenfill,
        text width=38mm,
    },
    derived/.style={
    	process,
    	draw=purple,
    	fill=purplefill,
    	minimum height=16mm
    },
    property/.style={
        process,
        draw=purple,
        fill=purplefill,
        text width=38mm,
        minimum height=24mm
    },
    output/.style={
        process,
        draw=green,
        fill=greenfill,
        text width=38mm
    },
    stage label/.style={
        font=\sffamily\bfseries\large,
        text=navy,
        anchor=west
    },
    note/.style={
        text=graytext,
        align=center
    }
]

\node[input] (dft) {\textbf{First-principles calculations}\\[1pt]
    QE or VASP};
\node[input, right=13mm of dft] (tb) {\textbf{Tight-binding models}\\[1pt]
    Slater-Koster, EDTB, or user-defined $\rightarrow$\code{PythTB}};
\node[input, right=13mm of tb] (phonopy) {\textbf{Force constants}\\[1pt]
    $\rightarrow$\code{phonopy}};

\node[process, below=16mm of dft] (proj) {\textbf{Project wavefunctions}\\[1pt]
    \textbf{onto PAO basis set}};
\node[process, below=8mm of proj] (ham) {\textbf{Construct PAO Hamiltonian}\\[1pt]
    $\mathcal{H}(\mathbf{k})$ and $\mathcal{H}(\mathbf{R})$};

\node[process, left=22mm of ham, draw=orange, fill=orangefill,
      text width=38mm] (uv) {\textbf{Compute Hubbard U+V parameters}\\[1pt]
    \code{PAOFLOW.ACBN0}};

\draw[flow] (dft) -- (proj);
\coordinate (A) at ($(phonopy.north |- dft.north)+(0,6mm)$);
\draw[flow] (dft.north) -- ([yshift=6mm]dft.north) -- (A) -- (phonopy.north);
\draw[flow] (proj) -- (ham);
\draw[flow] (tb.south) |- (ham.east);
\draw[flow] (ham.west) -- (uv.east);
\coordinate (cut1) at ($(uv.north)+(0,25mm)$);
\coordinate (cut2) at ($(uv.north)+(0,32mm)$);
\draw[flow_noarrow] (uv.north) -- (cut1);
\draw[flow] (cut2) |- (dft.west);

\node[note, text=orange, font=\sffamily\bfseries]
    at ($(uv.north)!0.5!(dft.west)+(0mm,-10mm)$) {\textbf{Self-consistent}\\\textbf{ACBN0 cycle}};

\draw[cycle] ($(dft.south)+(-0.2,0mm)$) -- ($(proj.north)+(-0.2,0mm)$);
\draw[cycle] ($(proj.south)+(-0.2,0mm)$) -- ($(ham.north)+(-0.2,0mm)$);
\draw[cycle] ($(ham.west)+(0,2mm)$) -- ($(uv.east)+(0,2mm)$);
\coordinate (cut3) at ($(uv.north)+(0.2,25mm)$);
\coordinate (cut4) at ($(uv.north)+(0.2,32mm)$);
\draw[cycle_noarrow] ($(uv.north)+(0.2,0mm)$) -- (cut3);
\draw[cycle] (cut4) |- ($(dft.west)+(0,-2mm)$);
\node[derived, below=17mm of ham,
text width=60mm] (momentum) {\textbf{Hamiltonian derivatives} \\ \textbf{(Required for some properties)} \\
	$\nabla_{\mathbf{k}}\mathcal{H}(\mathbf{k})$ and $\nabla^2_{\mathbf{k}}\mathcal{H}(\mathbf{k})$};
\draw[flow] (ham) -- (momentum);

\node[property, below=12mm of momentum, xshift=-64mm] (band) {
    \textbf{Electronic structure}\\[2pt]
    Band structure and Fermi surface\\[2pt]
    Quantum oscillations $\rightarrow$\code{pyskeaf}};

\node[property, right=2mm of band] (transport) {
    \textbf{Transport properties}\\[2pt]
    Semiclassical Boltzmann\\
    Kubo formalism\\
    Quantum transport\\[2pt]
    \code{PAOFLOW.Transport}};

\node[property, right=2mm of transport] (optic) {
    \textbf{Optical properties}\\[2pt]
    Dielectric permittivity\\[2pt]
    Optical \\conductivity ...};
    
\node[property, right=2mm of optic] (topology) {
	\textbf{Topology}\\[2pt]
	Topological invariants\\[2pt]
	$\rightarrow$\code{Z2Pack}};

\node[derived, below=73mm of phonopy] (phonon) {\textbf{Phonon} \\IR, Raman \\ Quasi-harmonic approx.};

\node[property, right=2mm of topology] (elph) {\textbf{Electron-phonon coupling} \\ Eliashberg function \\ ... };
\draw[flow] (phonopy) -- (phonon);
\draw[flow] (phonon) -- (elph);

\coordinate (propertybus) at ($(momentum.south)+(0,-6mm)$);
\draw[flow_noarrow] (momentum.south) -- (propertybus);
\draw[flow] (propertybus) -| (band.north);
\draw[flow] (propertybus) -| (transport.north);
\draw[flow] (propertybus) -| (optic.north);
\draw[flow] (propertybus) -| (topology.north);
\draw[flow] (propertybus) -| ($(elph.north)+(-0.1mm,0)$);
\node[output, below=16mm of optic] (plot) {
    \textbf{Plot and visualization}\\[1pt]
    \code{PAOFLOW.GPAO}};
\node[output, left=10mm of plot] (processing) {
   	\textbf{Further processing}\\[1pt]
   	Band unfolding \\ Site-, orbital-projection};

\coordinate (plotbus) at ($(optic.south)+(0,-6mm)$);
\draw[flow_noarrow] (band.south) |- (plotbus);
\draw[flow_noarrow] (transport.south) |- (plotbus);
\draw[flow_noarrow] (topology.south) |- (plotbus);
\draw[flow_noarrow] (optic.south) |- (plotbus);
\draw[flow_noarrow] (elph.south) |- (plotbus);
\draw[flow] (plotbus) -- (plot.north);
\draw[flow] (processing.east) -- (plot.west);
\draw[flow] (band.south) |- (processing.west);

\begin{scope}[on background layer]
    \node[fit=(dft)(tb)(phonopy), fill=stagefill, draw=grayline,
          rounded corners=4pt, inner sep=7mm] (stage1) {};
    \node[fit=(uv)(proj)(ham), fill=stagefill, draw=grayline,
          rounded corners=4pt, inner sep=7mm] (stage2) {};
    \node[fit=(band)(transport)(topology)(phonon)(momentum), fill=stagefill, draw=grayline,
          rounded corners=4pt, inner sep=7mm] (stage3) {};
    \node[fit=(plot)(processing), fill=stagefill, draw=grayline,
          rounded corners=4pt, inner sep=7mm] (stage4) {};
\end{scope}

\node[stage label] at ($(stage1.north west)+(2mm,-3mm)$) {1\quad Inputs};
\node[stage label] at ($(stage2.north west)+(2mm,-3mm)$) {2\quad Model construction};
\node[stage label] at ($(stage3.north west)+(2mm,-3mm)$) {3\quad Property calculations};
\node[stage label] at ($(stage4.north west)+(2mm,-3mm)$) {4\quad Outputs};

\end{tikzpicture}}
    \caption{Overview of the \paoflow{} workflow. Starting from first-principles calculations or tight-binding models, the PAO Hamiltonian is constructed and used to compute electronic, transport, optical, topological, and vibrational properties.}
    \label{fig:flowchart}
\end{figure*}

\section{The \paoflow{} fundamentals}
\label{sec:paoflow}
The \paoflow{} methodology is based on the construction of an accurate tight-binding Hamiltonian directly from first principles electronic structure calculations. Rather than fitting empirical parameters, the Kohn-Sham Bloch states obtained from a plane-wave DFT calculation are projected onto a localized basis, yielding a compact Hamiltonian that faithfully reproduces the selected electronic bands~\cite{Agapito2016_1,Agapito2016_2,DAmico2016}. 

\subsection{Projection onto the PAO basis}
\label{sec:pao_projection}
The construction of the PAO Hamiltonian starts from the self-consistent Kohn-Sham Bloch eigenstates $|\psi_n\rangle$ obtained using a plane-wave DFT code. These states are projected onto a localized basis of $M$ orthonormal orbitals $|\phi_\alpha\rangle$, which can be pseudoatomic orbitals or other localized basis spanning the subspace $\mathcal{A}$. The corresponding projection operator is:

\begin{equation}
\hat{P}
=
\sum_{\alpha}
|\phi_\alpha\rangle
\langle\phi_\alpha|,
\end{equation}

\noindent and the projected Bloch states:

\begin{equation}
|B_n\rangle
=
\hat{P}
|\psi_n\rangle
\end{equation}

\noindent are represented by the coefficients $ B_{\alpha n} = \langle\phi_\alpha|\psi_n\rangle$. 

The quality of the PAO representation is quantified by the projectability of the band:

\begin{equation}
p_n
=
\langle\psi_n|\hat{P}|\psi_n\rangle
=
\sum_{\alpha}
|B_{\alpha n}|^2,
\label{eq:proj}
\end{equation}

\noindent which measures how accurately a given Bloch state is represented within the chosen basis. A value of $p_n=1$ corresponds to an exact representation, whereas low projectability indicates that the basis is insufficient to describe the corresponding state. In practice, only states with sufficiently high projectability (typically $p_n\gtrsim0.9$) are retained for the Hamiltonian construction, while poorly represented states are discarded~\cite{Agapito2016_1}.



\subsection{Hamiltonian construction}
\label{sec:pao_ham}
After selecting the $N$ well-represented Bloch states, the normalized projection vectors:
\begin{equation}
|A_n\rangle
=
\frac{|B_n\rangle}{\sqrt{p_n}}
\end{equation}
\noindent are assembled into the projection matrix $A$, from which the PAO Hamiltonian is constructed as~\cite{Agapito2016_1}:
\begin{equation}
\bar{H}
=
A E A^\dagger,
\label{eq:Hbar}
\end{equation}
\noindent where $E=\mathrm{diag}(\varepsilon_1,\ldots,\varepsilon_N)$ contains the corresponding Kohn-Sham eigenvalues.

Because only a subset of the PAO basis is used, the projected Hamiltonian formally contains an artificial null space of dimension $M-N$. This subspace does not correspond to physical electronic states and is removed by introducing a shift projector:
\begin{equation}
\bar{H}
\rightarrow
\bar{H}
+
\kappa Q_{\mathcal N},
\qquad
Q_{\mathcal N}
=
I_M
-
A(A^\dagger A)^{-1}A^\dagger,
\label{eq:shift}
\end{equation}
\noindent which moves the null-space eigenvalues to an energy $\kappa$ above the energy window of interest without affecting the physical eigenstates. In practice, $Q_{\mathcal{N}}$ may be approximated as $I_M-AA^\dagger$ when only very high-projectability states are considered.


The resulting PAO Hamiltonian is initially obtained on the DFT $\mathbf{k}$-mesh used in the original first principles calculation. It is then transformed to its real-space representation:
\begin{equation}
H_{\alpha\beta}(\mathbf{R})
=
\frac{1}{N_k}
\sum_{\mathbf{k}}
e^{-i\mathbf{k}\cdot\mathbf{R}}
H_{\alpha\beta}(\mathbf{k}),
\label{eq:realspace}
\end{equation}
\noindent where $N_k$ is the number of $\mathbf{k}$-points in the Brillouin zone. The Hamiltonian at an arbitrary wave vector is then reconstructed by:
\begin{equation}
H_{\alpha\beta}(\mathbf{k})
=
\sum_{\mathbf{R}}
e^{i\mathbf{k}\cdot\mathbf{R}}
H_{\alpha\beta}(\mathbf{R}),
\label{eq:fourier}
\end{equation}

\noindent allowing electronic bands and wave functions to be evaluated on arbitrarily dense $\mathbf{k}$-meshes. These interpolated quantities form the basis of all subsequent electronic, optical, transport, and topological properties calculations.

 
\subsection{DFT interfaces and basis presets}
\label{sec:theory:intbasis}
\begin{table*}[!htb]
	\begin{tabular*}{\textwidth}{@{\extracolsep{\fill}}ccccc}
		\hline
		Method & Function & \code{configuration} & Basis & DFT code \\ \hline
		\textbf{1} & \makecell{\code{PAOFLOW.}\\ \code{read\_atomic\_proj\_QE()}} & None & PP & \multirow{4}{*}{QE} \\ \cline{1-4}
		\textbf{2} & \multirow{5}{*}{\code{PAOFLOW.projections()}} & \code{`minimal'} & PP & \\ \cline{1-1} \cline{3-4}
		\multirow{2}{*}{\textbf{3}} & & \makecell{\code{`standard'} \\ or \code{ `extended'} or \\\code{\{user-defined Python dict\}}} & \makecell{\code{PAOFLOW.basis\_gen} \\ from PP (recommended) \\ or PAOFLOW/BASIS/} & \\ \cline{3-5}
         & & \code{\{user-defined Python dict\}} & PAOFLOW/BASIS/ & VASP \\ \hline
	\end{tabular*}
    \caption{Summary of the different basis configuration schemes available in \paoflow{}.}
    \label{tab:basis}
\end{table*}

\paoflow{} currently interfaces with two plane-wave DFT codes: Quantum
ESPRESSO (QE)~\cite{Giannozzi2009,Giannozzi2017}
and VASP~\cite{Kresse1996,Kresse1999}.
For QE, the projection amplitudes
$B_{\alpha n\mathbf{k}} = \langle\phi_\alpha|\psi_{n\mathbf{k}}\rangle$
can be obtained in three ways. \textbf{1.} When projection is already performed by QE, it can be read directly from the output of \code{projwfc.x}. \textbf{2.} The wavefunctions produced by \code{pw.x} are projected onto pseudo-atomic wavefunctions stored in the pseudopotential (PP), giving identical projection amplitudes as Method \textbf{1}. \textbf{3.} The wavefunctions are projected onto the all-electron PAO basis distributed with \paoflow{} (located in the PAOFLOW/BASIS/ folder), or generated on the fly with \code{PAOFLOW.basis\_gen} from PP files, making the basis fully specific to the PP used in the simulation. The latter is essential for fully relativistic, j-resolved pseudopotentials.

For Method \textbf{2} and \textbf{3}, \paoflow{} introduces a hierarchical \emph{basis preset} system to facilitate the selection process, i.e. \code{projections(configuration=\ldots)} where \code{configuration} can take values of
\begin{itemize}
	\item \code{`minimal'}: use Method \textbf{2}.
	\item \code{`standard'}: use Method \textbf{3}, all-electron PAO basis with
	`minimal' valence configuration augmented with the next missing
	angular-momentum channel (e.g. `3D' for
	Si) and an additional radial function for each occupied shell (e.g. `4S', `4P' for
	Si).
	\item \code{`extended'}: use Method \textbf{3}, further augmenting the `standard' set with
	a generous set of rule-based polarization shells. Benchmarks on representative elements show that the extended
	preset can triple the number of states with projectability
	$p_n > 0.95$~(Eq.~\eqref{eq:proj}) for conduction bands without degrading the valence description.
\end{itemize}

A suitable basis set should contain enough orbitals to accurately reproduce the states in the desired energy range, while avoiding unnecessary basis functions. Although Methods \textbf{1} and \textbf{2} reproduce the valence states with high fidelity, the basis set provided in the pseudopotential contains only a limited number of orbitals and therefore often provides an inadequate description of higher-energy conduction states. These states are important for calculating properties such as the dielectric permittivity. In such cases, Method \textbf{3} is preferable, as the enlarged basis allows a greater number of conduction states to be accurately reproduced. Additionally, fully user-defined configurations are supported, e.g. \code{configuration=\{`Si':[`3S',`3P',`3D',`4S'],\\`O':[`2S',`2P',`3S',`3P']\}}, as well as a mixed scheme allowing a curated preset on some species while hand-picking orbitals on others, e.g., \code{configuration=\{`Ga': `standard', `As': [`4S', `4P', `3D']\}}.

\begin{figure*}[!t]
    \centering
    \begin{minipage}{0.49\textwidth}
        \centering
        \includegraphics[width=\linewidth,height=0.32\textheight,keepaspectratio]{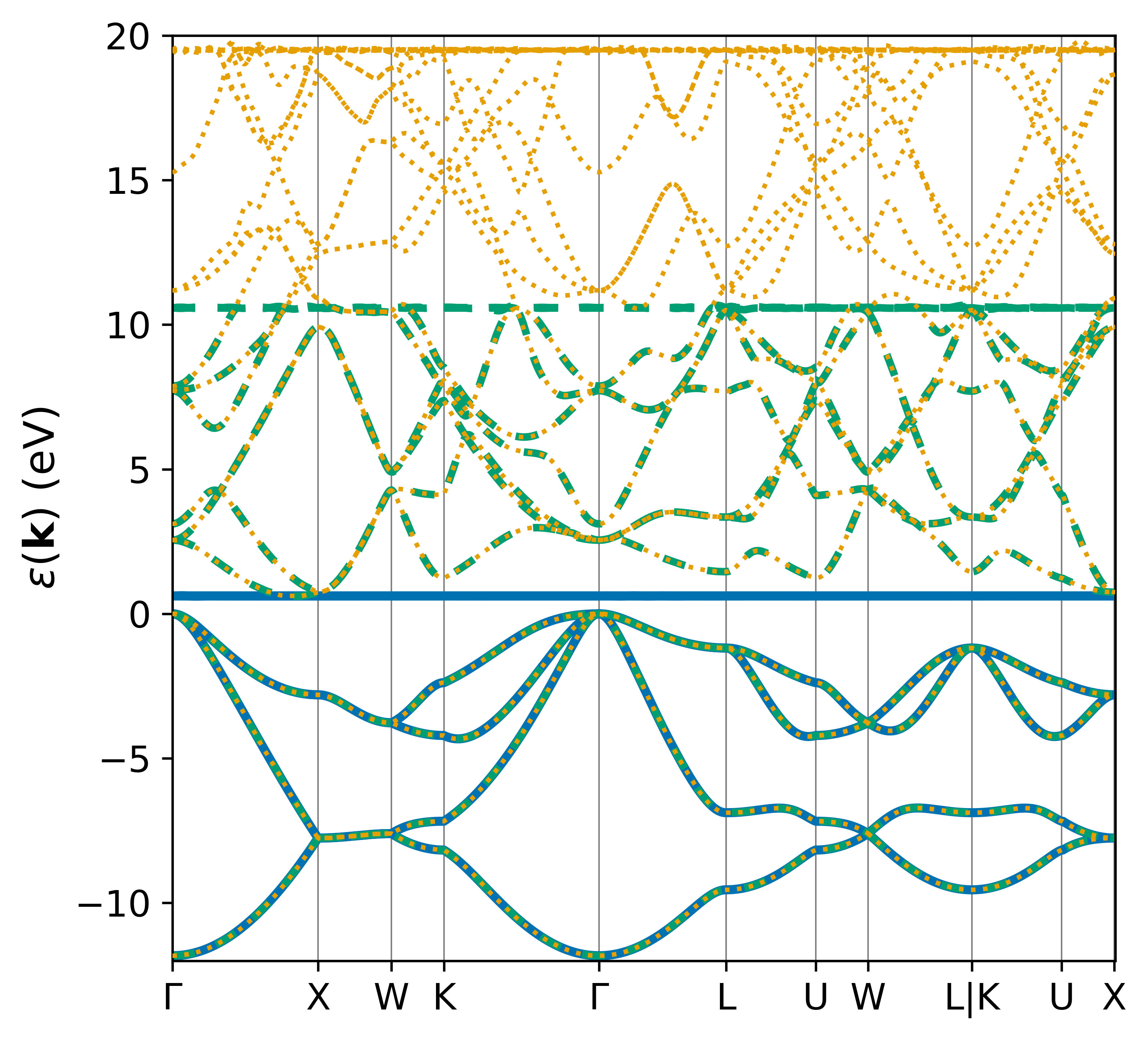}
    \end{minipage}\hfill
    \begin{minipage}{0.49\textwidth}
        \centering
        \includegraphics[width=\linewidth,height=0.32\textheight,keepaspectratio]{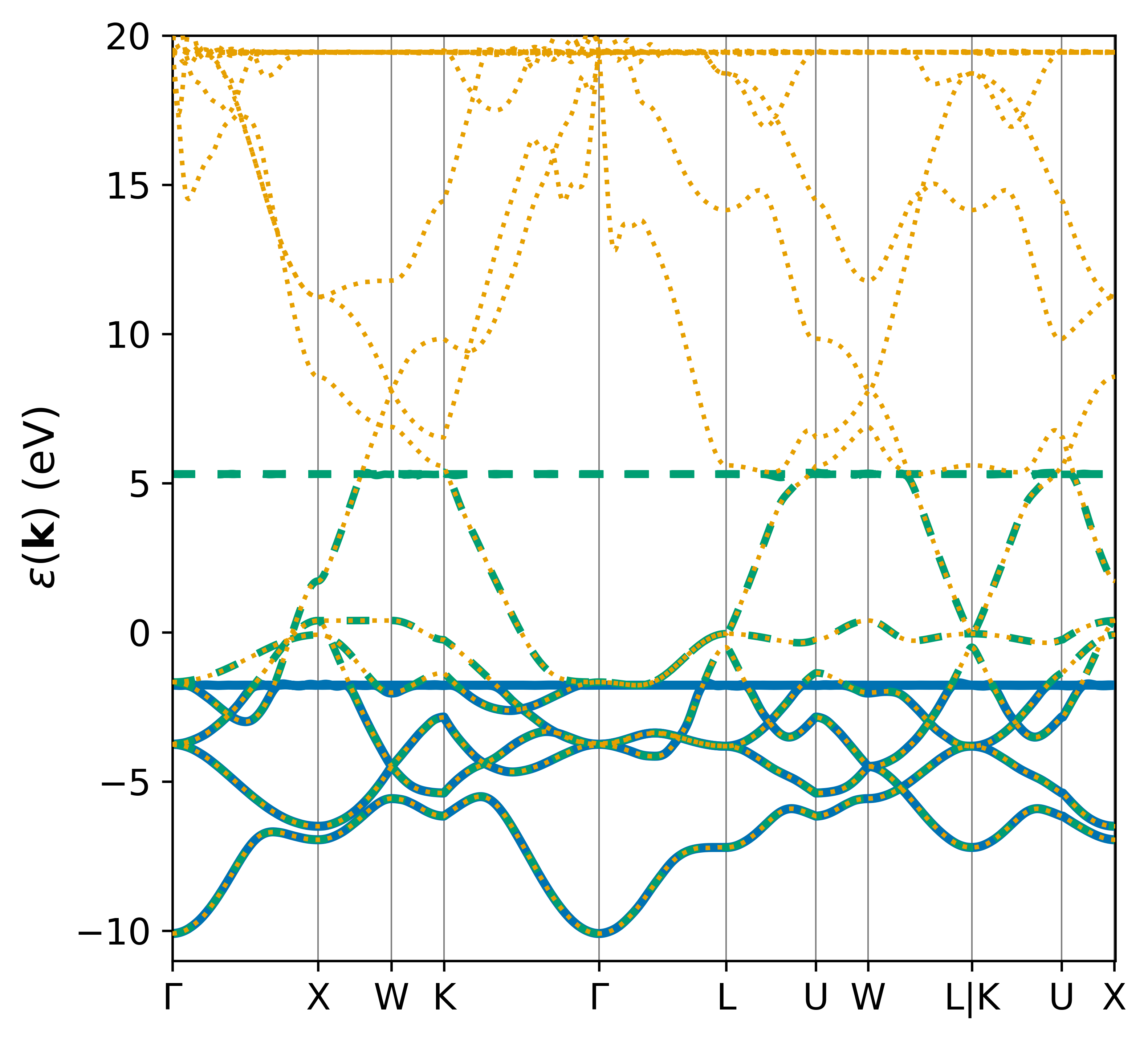}
    \end{minipage}
    \caption{Band structures obtained with the minimal (blue)/standard (green)/extended (yellow) PAO basis: (left) Si, (right) Pt. The horizontal lines correspond to the null-space eigenvalues, pushed out of the region of high-projectability (see Sec. \ref{sec:pao_ham}).}
    \label{fig:basis}
\end{figure*}

In Fig.~\ref{fig:basis} we compare the band structures of silicon and platinum obtained with the minimal, standard, and extended PAO presets. As the basis is enlarged, the PAO-projected Hamiltonian not only preserves the low-energy manifold but extends the energy window over which the DFT bands are faithfully reproduced, capturing higher-lying conduction states (Si) and more strongly hybridized states (Pt) with high accuracy. 

For VASP, the projection should be carried out only with Method \textbf{3} with user-defined basis configuration. The wavefunction from \code{WAVECAR} are projected onto the all-electron PAO basis distributed with \paoflow{}. The different basis configuration schemes are summarized in Table \ref{tab:basis}.

\subsection{Hamiltonian derivatives}
\label{sec:theory:derivatives}

Many of the response properties evaluated by \paoflow{} -- group
velocities, the dielectric function, Berry curvatures, anomalous Hall
and spin Hall conductivities, Boltzmann transport
coefficients -- require the momentum derivative
$\partial H(\vk)/\partial k_l$ of the PAO Hamiltonian.  Because
$\bar{H}(\vR)$ is known analytically on a uniform real-space grid -- it
is the inverse Fourier transform of the Bloch Hamiltonian on the DFT
$\vk$-mesh, Eq.~\eqref{eq:fourier} -- differentiation reduces to a
multiplication by the position operator in real space:
\begin{equation}
  \frac{\partial H_{\alpha\beta}(\vk)}{\partial k_l}
   = i\,\sum_{\vR} R_l\,
     e^{i\vk\cdot\vR}\,H_{\alpha\beta}(\vR),
  \label{eq:dHdk}
\end{equation}
where $R_l$ is the $l$-th Cartesian component of the lattice vector
$\vR$.  In practice the sum in Eq.~\eqref{eq:dHdk} is carried out via a
pair of fast Fourier transforms on the same mesh used to build
$H(\vk)$: an inverse fast Fourier transform (FFT) brings the Hamiltonian to real space, the
result is multiplied pointwise by $R_l$, and a forward FFT returns the
gradient on the dense interpolation mesh.  The procedure inherits the
$\mathcal{O}(N_{\vk}\log N_{\vk})$ scaling of the FFT and is performed
in place over the distributed orbital index.

The PAO basis is orthonormal by construction, so no overlap-matrix
contribution enters the gradient.  However, Eq.~\eqref{eq:dHdk} is
written in the so-called \emph{periodic} (or lattice) gauge, in which
the Bloch sum carries phase factors $e^{i\vk\cdot\vR}$ tied solely to
the Bravais lattice vectors $\vR$ and the intracell positions
$\boldsymbol{\tau}_\alpha$ of the basis orbitals do not appear.  Most
physical observables, on the other hand, require derivatives evaluated
in the \emph{atomic} (or tight-binding) gauge, where the Bloch phases
include the orbital centers, $e^{i\vk\cdot(\vR+\boldsymbol{\tau}_\beta
-\boldsymbol{\tau}_\alpha)}$ (see, e.g., the discussion of the two
conventions in Refs.~\cite{Vanderbilt2018,Tsirkin2021}). The change of
gauge introduces an additive correction which, for the first and second derivatives, is given by:
\begin{equation}
    \centering
    \begin{split}
      \frac{\partial H_{\alpha\beta}(\vk)}{\partial k_l}
       \longrightarrow
       \frac{\partial H_{\alpha\beta}(\vk)}{\partial k_l}
       + i\,H_{\alpha\beta}(\vk)\,D^{\alpha\beta}_l,
  \end{split}
  \label{eq:dHdk_Dnm}
\end{equation}
\begin{equation}
  \centering
  \begin{aligned}
    \frac{\partial^2 H_{\alpha\beta}(\vk)}{\partial k_l\partial k_m}
    \longrightarrow
    \frac{\partial^2 H_{\alpha\beta}(\vk)}{\partial k_l\partial k_m}
    & + H_{\alpha\beta}(\vk)\,D^{\alpha\beta}_l\,D^{\alpha\beta}_m \\
    &\hspace{-6.5em} + i\,D^{\alpha\beta}_l\,\frac{\partial H_{\alpha\beta}(\vk)}{\partial k_m}
      + i\,D^{\alpha\beta}_m\,\frac{\partial H_{\alpha\beta}(\vk)}{\partial k_l}
  \end{aligned}
  \label{eq:d2Hdk2_Dnm}
\end{equation}
where $D^{\alpha\beta}_l = \tau_{\beta,l}-\tau_{\alpha,l}$ is the Cartesian component of the
intracell-position difference between the two orbital centers.  This
term vanishes only for a single-atom basis, where all
orbitals share the same center; for any multi-atom unit cell it must be
retained, and is precomputed once from the orbital positions.
Equations~\eqref{eq:dHdk} to ~\eqref{eq:d2Hdk2_Dnm} provide an analytic,
finite-difference-free gradient that is consistent with the
Hellmann-Feynman theorem and is used throughout the modules
described in the following sections.

\subsection{Spin-orbit formalism}
\label{sec:theory:spinorb}
Spin-orbit coupling (SOC) is a fundamental ingredient in
condensed-matter physics, and several properties can only be properly
described using fully relativistic calculations. However, DFT
calculations including SOC can be computationally  extremely demanding.
\paoflow{} provides the possibility of introducing SOC into a
PAO Hamiltonian, offering a computationally efficient approach. 
We start from the scalar-relativistic PAO Hamiltonian,
\begin{equation}
H_0=\sum_{ij}\sum_{\mu\nu}\sum_{\sigma}t_{ij}^{\mu\nu}
c^\dagger_{i\mu\sigma}c^{}_{j\nu\sigma},
\label{hopping}
\end{equation}
where $c^\dagger_{i\mu\sigma} (c^{}_{i\mu\sigma})$ denotes the
creation (annihilation) operator for an electron with spin projection
$\sigma$ at atomic site $i$ and orbital $\mu$. Here, the labels $i$
($j$), $\mu$ ($\nu$), and $\sigma$ denote atomic sites, orbitals, and
spin projections, respectively. The quantity $t_{ij}^{\mu\nu}$ denotes
the corresponding hopping matrix element of the PAO Hamiltonian. The
atomic SOC is introduced through the term
\begin{equation}
H_\mathrm{SOC}=\sum_i\sum_{\mu\nu}\sum_{\sigma\sigma'}\xi^{\mu\nu}_i
\bra{i\mu\sigma}\bm{L}\cdot\bm{S}\ket{i\nu\sigma'} c^\dagger_{i\mu\sigma}c^{}_{i\nu\sigma'},
\label{LdotS}
\end{equation}
where $\bm{L}$ and $\bm{S}$ are the orbital and spin angular momentum
operators, respectively. The SOC strength $\xi^{\mu\nu}_i$ can be obtained
from a fit to relativistic DFT calculations. This methodology has been
successfully applied systems ranging from topological to magnetic materials~\cite{costa18,Costa19,costa20,cardias25,Dugato25}.

\subsection{Optical and dielectric response}
\label{sec:theory:optical}

The frequency-dependent dielectric tensor is evaluated in the
independent-particle approximation via the Kubo-Greenwood
formula~\cite{Kubo1957,Greenwood1958}.
The complex dielectric function $\tilde{\varepsilon}(\omega) = \varepsilon_1(\omega) + i\varepsilon_2(\omega)$ is calculated via:
%
\begin{equation}
    \begin{split}
        \tilde{\varepsilon}_{\alpha\beta}(\omega) & =
\delta_{\alpha\beta} + \frac{4\pi e^2}{\Omega N_k}
\sum_{n,\mathbf{k}}
\frac{d f(E_{\mathbf{k},n})}{d E_{\mathbf{k},n}}
\frac{\hat{M}_{\alpha,\beta}}
{\omega^2+i\eta\omega}+\frac{8\pi e^2}{\Omega N_k}\times\\ &
\sum_{m\neq n,\mathbf{k}}
\frac{\hat{M}_{\alpha,\beta}}
{E_{\mathbf{k},m}-E_{\mathbf{k},n}}
\frac{f(E_{\mathbf{k},n})}
{(\omega_{\mathbf{k},m}-\omega_{\mathbf{k},n})^2
-\omega^2-i\Gamma\omega}.
    \end{split}
\end{equation}
where $\Omega$ is volume of the unit cell, $N_k$ the number of k-points and $\hat{M}_{\alpha,\beta}=\langle n\mathbf{k} | \hat{v}_\alpha | m\mathbf{k}\rangle\langle m\mathbf{k} | \hat{v}_\beta  | n\mathbf{k}\rangle$ for $\hat{v}_\alpha$ being the velocity operator (Eq. \eqref{eq:dHdk_Dnm}). $\Gamma$ and $\eta$, as user inputs, control the broadening of the inter-band and intra-band transition respectively. For the Fermi-Dirac distribution $f(E)$ and its derivative, adaptive smearing with the local band velocity~\cite{Yates2007} is available alongside fixed Gaussian and Methfessel-Paxton smearing.
From the dielectric tensor, a few useful quantities can be derived, among which:

\noindent
the plasmon frequency estimated from the $f$-sum rule: \begin{equation}\omega_p^2 = \frac{2}{\pi}\int_0^\infty\omega\,\varepsilon_2(\omega)\,d\omega,\end{equation}
the optical conductivity:
\begin{equation}\tilde{\sigma}(\omega) = -i\,\varepsilon_0\,\omega\,
                  \bigl(\tilde{\varepsilon}(\omega) - 1\bigr),\end{equation}
the electron energy-loss spectrum (EELS):
\begin{equation}L(\omega) = -\mathrm{Im}\,\left(\frac{1}{\tilde{\varepsilon}(\omega)}\right),\end{equation}
and the complex refractive index $\tilde n = n + i\kappa$ which satisfies $\tilde n^2 = \tilde{\varepsilon}$, with 
\begin{equation}
n      
= \sqrt{(|\tilde{\varepsilon}| + \varepsilon_1)/2}, \quad
\kappa = \sqrt{(|\tilde{\varepsilon}| - \varepsilon_1)/2}.
\end{equation}

\subsection{Nonlocal pseudopotential velocity correction}
\label{sec:theory:nlv}

When utilizing norm-conserving pseudopotentials, calculating the electronic velocity via a simple analytical derivative of the Hamiltonian is insufficient. 
In fact, the PAO-based
velocity operator $\hat{p}_\alpha^{\mathrm{PAO}} =
m_e\hbar^{-1}\partial H^{\mathrm{PAO}}/\partial k_\alpha$ inherently assumes a local potential energy landscape. However, physical norm-conserving pseudopotentials frequently introduce a spatially non-local, separable component to accurately model core-valence electron interactions. This non-local potential is typically expressed in the Kleinman-Bylander form
$\hat{V}_{\mathrm{NL}} = \sum_{ij}|\beta_i\rangle D_{ij}\langle\beta_j|$,
and the gauge-invariant momentum operator acquires an additional commutator
contribution~\cite{Read1991,Hyber-Louie1987,DelSole1996}:
\begin{equation}
  \hat{\mathbf{p}}^{\mathrm{full}} = \hat{\mathbf{p}} +
  \frac{m_e}{i\hbar}\bigl[\hat{V}_{\mathrm{NL}},\hat{\mathbf{r}}\bigr].
  \label{eq:p_full}
\end{equation}

For a separable Kleinman-Bylander
pseudopotential~\cite{Kleinman1982}, the correction in the PAO-projected basis
is
\begin{equation}
\begin{split}
  \Delta p_{\mu\nu,\alpha}(\mathbf{k}) =
    \frac{m_e}{i\hbar} \sum_{ij} D_{ij}
    \Bigl[
      \langle\beta_i|\varphi_{\mu\mathbf{k}}\rangle\,
      \langle r_\alpha\beta_j|\varphi_{\nu\mathbf{k}}\rangle^{*}
    - \\ \langle r_\alpha\beta_i|\varphi_{\mu\mathbf{k}}\rangle\,
      \langle\beta_j|\varphi_{\nu\mathbf{k}}\rangle^{*}
    \Bigr],
  \label{eq:Delta_p}
  \end{split}
\end{equation}
where $\langle\beta_i|\varphi_{\mu\mathbf{k}}\rangle$ are the projector
overlaps stored in the DFT output and the position-weighted overlaps
$\langle r_\alpha\beta_j|\varphi_{\nu\mathbf{k}}\rangle$ are constructed from
the radial integrals of the projector functions.

The nonlocal velocity correction alters off-diagonal matrix elements, making it crucial for BZ sums in optical absorption. However, it vanishes for diagonal elements and leaves properties derived from the Kubo formalism practically unchanged because the dominant contributions are weighted by energy denominators, and the correction mainly redistributes oscillator strength at higher energies.

\section{ACBN0: A Pseudohybrid Hubbard density functional}
\label{sec:acbn0:acbn0}

\begin{figure*}[!htb]
    \centering
    \begin{minipage}[c]{0.50\textwidth}
        \centering
        \includegraphics[height=0.32\textheight]{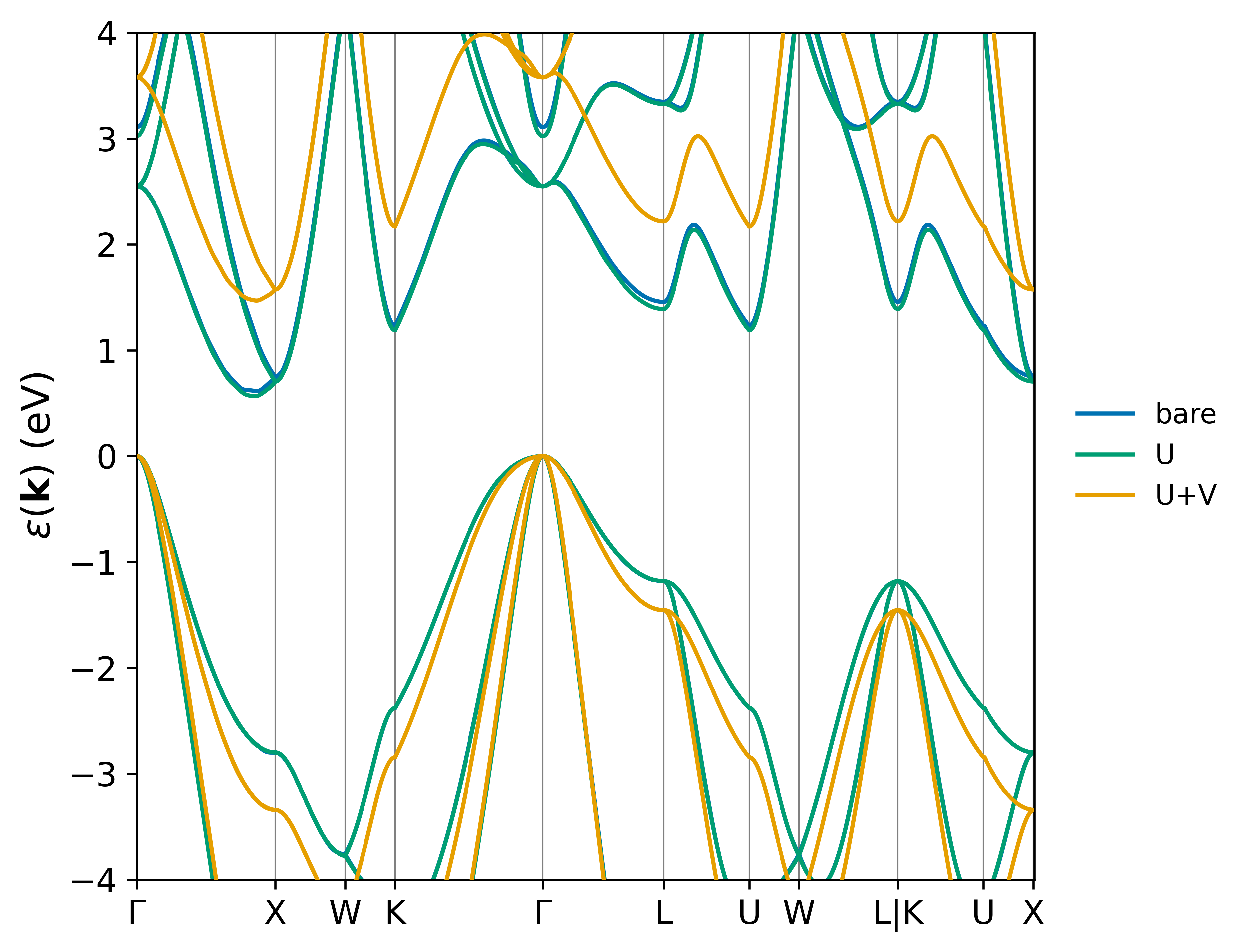}
    \end{minipage}\hfill
    \begin{minipage}[c]{0.46\textwidth}
        \centering
        \includegraphics[height=0.32\textheight]{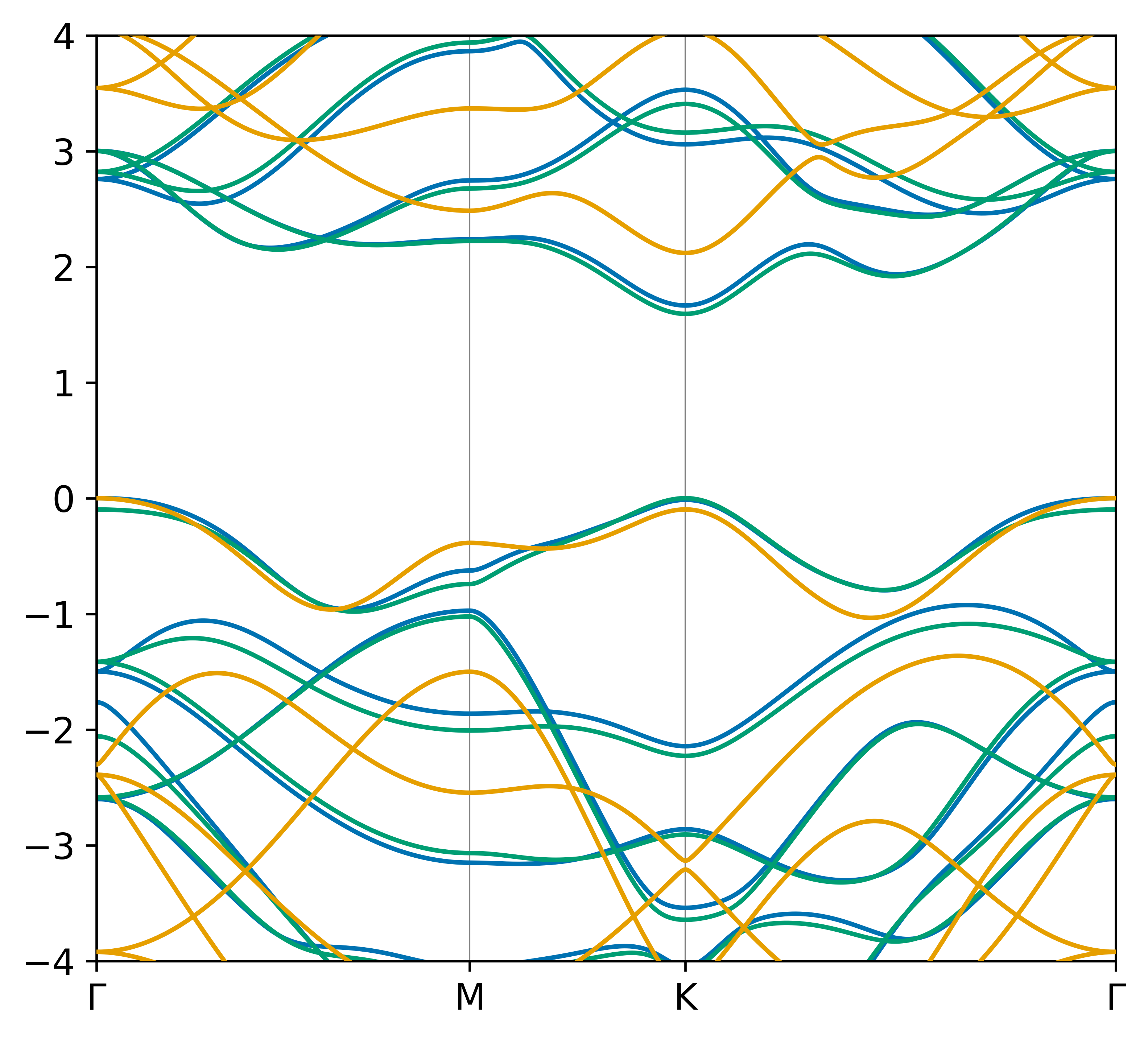}
    \end{minipage}
    \caption{Band structures for silicon (left) and MoS$_2$ (right) comparing the uncorrected DFT calculations with the DFT+U and DFT+U+V from eACBN0. Values for silicon are $U(p) = 2.7, ~ V(s{-}s) = 1.39, ~V(s{-}p) = 0.73, ~{\rm and} ~ V(p{-}p) = 1.96$ and $E_{\rm g} = 1.46$ in agreement with~\cite{Lee2020eACBN0},  and for MoS$_2$ $U({\rm S}-3p) = 1.77,~U({\rm Mo}-4d) = 3.09, ~ V(d{-}p) = 1.19$, and $E_{\rm g} = 2.22$ in good agreement with hybrid functionals and experiments~\cite{perdew2024}. All values are given in eV. }
    \label{fig:acbn0}
\end{figure*}

Standard LDA and GGA exchange-correlation functionals notoriously
underestimate the band gaps of insulators and semiconductors and fail to
capture the electronic structure of strongly correlated systems such as
transition-metal oxides (TMOs).  The DFT$+U$ correction of Liechtenstein
and Anisimov~\cite{Liechtenstein1995,Dudarev1998} partially
remedies this by augmenting the DFT energy with an explicit on-site
Hubbard-like term:
\begin{equation}
  E^{\mathrm{DFT}+U} = E^{\mathrm{DFT}} + E_U,
\end{equation}
where $E_U$ depends on the effective on-site Coulomb ($\bar{U}$) and
exchange ($\bar{J}$) parameters.  In practice these parameters are
treated as empirical inputs, which limits the predictive power and
transferability of the method.

The pseudohybrid Hubbard density functional ACBN0 introduced by L. Agapito, S. Curtarolo and M. Buongiorno Nardelli~\cite{Agapito2015ACBN0}
eliminates this empiricism by computing $\bar{U}$ and $\bar{J}$
self-consistently from first principles.  The key idea is to evaluate
the on-site Hartree--Fock Coulomb and exchange energies directly from
the electron repulsion integrals (ERIs) of the localized atomic-orbital
set $\{m\}$ centered on the Hubbard atom of interest:
\begin{equation}
\begin{split}
  E_{\mathrm{HF}}^{\{m\}} = \tfrac{1}{2}\sum_{\{m\},\sigma}
  \bar{P}^{\sigma}_{mm'}\bar{P}^{\sigma}_{m''m'''} \\
  \times\bigl[2(mm'|m''m''') - (mm'''|m''m')\bigr],
\end{split}
\end{equation}
where the ERIs
\begin{equation}
    \centering
    \begin{split}
      & \qty(mm'|m''m''') = \\
      & \quad \int\!\dd \vb{r}_1\,\dd \vb{r}_2\;
      \phi_m(\vb{r}_1)\phi_{m'}(\vb{r}_1) 
      \ V\,\phi_{m''}(\vb{r}_2)\phi_{m'''}(\vb{r}_2)
    \end{split}
\end{equation}
are evaluated using pseudo-atomic orbitals (PAOs) expressed in a
minimal PAO-3G Gaussian basis.  The crucial ingredient is the
\emph{renormalized} density matrix $\bar{P}^{\sigma}_{\mu\nu}$, in
which the occupation of each Kohn--Sham state $\psi^{\sigma}_i$ is
replaced by its Mulliken charge projected onto the set $\{\bar{m}\}$
of all orbitals in the unit cell sharing the same angular-momentum
quantum numbers as $\{m\}$:
\begin{equation}
  \bar{N}^{\sigma}_{\psi_i}
  = \sum_{\mu\in\{\bar{m}\},\nu}
    c^{\sigma}_{\mu i} S_{\mu\nu} c^{\sigma}_{\nu i}.
\end{equation}
Replacing the full occupancy $N^{\sigma}_{\psi_i}=1$ with
$\bar{N}^{\sigma}_{\psi_i}\leq 1$ effectively renormalizes the density
matrix and introduces an implicit, mean-field screening of the bare
Coulomb interaction without requiring an explicit model for $V^{ee}$.
Comparing $E_{\mathrm{HF}}^{\{m\}}$ with the Dudarev form of $E_U$
then yields $\bar{U}$ and $\bar{J}$ as explicit density functionals,
satisfying the Hohenberg--Kohn requirement.

In the context of plane-wave DFT, the PAO-projected Kohn--Sham
wave functions needed to construct the renormalized density matrices
are obtained via the same PAO projection scheme that underpins
\paoflow{}~\cite{Agapito2016_1,paoflow1}.  The ERIs are
computed using a Gaussian-type representation of the PAO radial
functions, making the evaluation efficient with standard
quantum-chemistry routines.

At each cycle of the self-consistent ACBN0 procedure the Kohn--Sham eigenvalues from the current DFT$+U$
calculation are used to recompute $\bar{U}$ and $\bar{J}$, which then
seed the next DFT$+U$ step, until $|U^{(n)}-U^{(n-1)}|$ is smaller than a user-defined threshold.
$\bar{U}$ and $\bar{J}$ can be computed on an arbitrary number of orbitals on any number of sites simultaneously, making the method extremely versatile.  The result is a
parameter-free correction that reduces to standard LDA/GGA when
localization is absent, gives accurate band gaps for TMOs at a cost
only marginally above plain DFT$+U$, and satisfies all five criteria
outlined by Pickett for a well-defined Hubbard
functional~\cite{Pickett1998}.

\subsection{Extended ACBN0: self-consistent on-site $U$ and intersite $V$}
\label{sec:acbn0:eacbn0}

For systems with sizeable hybridization between neighbouring Hubbard
sites --- e.g.\ transition-metal oxides with significant
$d$--$p$ covalency, or layered chalcogenides where on-site corrections
alone are known to bias structural and electronic
properties~\cite{CampoJr2010} --- the on-site $U$ is not
sufficient and an \emph{intersite} Hubbard term $V$ must be added to
the energy functional.  The DFT$+U+V$ form of Campo and
Cococcioni~\cite{CampoJr2010},
\begin{equation}
  E^{\mathrm{DFT}+U+V} = E^{\mathrm{DFT}} + E_U + E_V,
\end{equation}
augments $E_U$ with a pairwise contribution
\begin{equation}
  E_V = -\tfrac{1}{2}\sum_{I\neq J,\sigma}
        V^{IJ} \sum_{ij}
        n^{IJ\sigma}_{ij}\,n^{JI\sigma}_{ji},
\end{equation}
where $n^{IJ\sigma}_{ij}$ is the generalized occupation matrix coupling
orbital $i$ on site $I$ to orbital $j$ on site $J$.  Just as for the
on-site $U$, in conventional DFT$+U+V$ the parameters $V^{IJ}$ are
either fitted empirically or computed via a linear-response procedure,
which scales unfavourably with the number of inequivalent pairs.

To overcome this limitation we have extended the ACBN0 construction to
the intersite case following Lee and Son~\cite{Lee2020eACBN0}.  The
extended-ACBN0 (eACBN0) parameter for the pair $(I,J)$ separated by a
Bravais lattice vector $\mathbf{R}$ reads:
\begin{equation}
  \label{eq:eacbn0:V}
  V^{IJ}(\mathbf{R}) =
    \frac{N^{IJ}(\mathbf{R})}{D^{IJ}(\mathbf{R})},
\end{equation}
with numerator:
\begin{equation}
\begin{split}
  N^{IJ}(\mathbf{R}) = \tfrac{1}{2}\sum_{ikjl}
    \Bigl[\,
      P^{II}_{ik}\,P^{JJ}_{jl}
      - \\ \sum_{\sigma}
        P^{IJ\sigma}_{il}(\mathbf{R})\,P^{JI\sigma}_{jk}(-\mathbf{R})
    \,\Bigr] 
    \times\,
    (ik\,|\,jl)_{\mathbf{R}},
\end{split}
\end{equation}
and denominator obtained analogously from the bare occupations
$n^{II}$, $n^{JJ}$, $n^{IJ}(\mathbf{R})$.  Here
$P^{II}_{ik} = \sum_{\sigma}P^{II\sigma}_{ik}$ is the spin-summed
on-site renormalized density matrix entering the direct (Hartree-like)
contribution, while the exchange-like term is intrinsically same-spin.
The four-centre integrals
$(ik\,|\,jl)_{\mathbf{R}}$ couple PAO Gaussians centred on site $I$ in
the home cell to those centred on site $J$ in the image cell labelled
by $\mathbf{R}$, and are evaluated with the same minimal PAO-3G basis
already used for the on-site Hartree--Fock integrals.  The minimum-image
convention is applied to keep the support of the basis functions
overlapping the home cell, ensuring rapid decay of $V^{IJ}(\mathbf{R})$
with $|\mathbf{R}|$.

In practice, the user registers the relevant pairs either by giving an
explicit list of $(I,J,n_a,n_b,n_c)$ tuples or by specifying a
real-space cutoff together with the species pairs of interest.
\paoflow{} then drives a joint $U+V$ self-consistent loop in which, at
each iteration, the current values of $U$ and $V$ are used to perform
the underlying DFT$+U+V$ calculation and to project the resulting
Kohn--Sham states onto the PAO basis.  The PAO Hamiltonian is
reconstructed and the on-site parameters
$\bar{U}$ and $\bar{J}$ are recomputed as in ordinary ACBN0 while the
intersite $V^{IJ}(\mathbf{R})$ are evaluated for every registered pair
through Eq.~\eqref{eq:eacbn0:V} using the same Coulomb-integral kernel
employed for the on-site term.  Linear mixing is applied to both $U$
and $V$, and convergence is monitored on the largest element of
$\{|\Delta U|,|\Delta V|\}$.
Because the renormalized density matrices $P^{IJ\sigma}(\mathbf{R})$
that enter Eq.~\eqref{eq:eacbn0:V} are built directly from the PAOFLOW
real-space Hamiltonian $H(\mathbf{R})$, they inherit all of its
gauge-related improvements: in particular, the per-$\mathbf{k}$
projectability filter guarantees that low-projectability
bands cannot pollute the on-site or intersite occupations.  The
self-consistent eACBN0 thus retains the parameter-free,
density-functional character of ACBN0 while extending its applicability
to systems where intersite hybridization is non-negligible.

In Fig \ref{fig:acbn0} we show a comparison of band structures calculated without correction, with U only and U+V for Si and MoS$_2$.
\section{Environment-dependent tight-binding from PAO Hamiltonians}
\label{sec:SKTB}

\subsection{Slater-Koster representation}
\label{sec:SKTB:sk}

The PAO Hamiltonian constructed by \paoflow{} and a Slater-Koster (SK)
tight-binding (TB) model occupy the same Hilbert space: both are expressed in a
localized basis labeled by identical angular-momentum quantum numbers, and both
span a subspace of the same dimension. A PAO Hamiltonian may therefore serve as
an exact \emph{ab initio} reference for the construction of transferable TB
models, without the gauge ambiguity that attends parameterizations based on
maximally localized Wannier functions. This correspondence underlies the
environment-dependent tight-binding (EDTB) scheme of
Ref.~\cite{Marco2026EDTB}, in which the formalism, its validation against
first-principles band structures, and the origin of its transferability are
developed in detail. Here we summarize the elements of that scheme as
implemented in \paoflow{} and illustrate their application to a large moir\'e
superlattice.

In the SK framework~\cite{SlaterKoster1954}, the hopping matrix element between
orbital $\alpha$ on atom $i$ and orbital $\beta$ on atom $j$, connected by the
bond vector $\mathbf{d}_{ij}$, decomposes as
\begin{equation}
  H_{\alpha\beta}^{\mathrm{SK}}(\mathbf{d}_{ij})
  = \sum_{\mu} c_{\alpha\beta\mu}(\hat{\mathbf{d}}_{ij})\,
    V_{l_\alpha l_\beta \mu}^{(s)} ,
\label{eq:sk_hop}
\end{equation}
where $c_{\alpha\beta\mu}$ are direction-cosine factors determined by
$\hat{\mathbf{d}}_{ij} = \mathbf{d}_{ij}/|\mathbf{d}_{ij}|$,
$\mu \in \{\sigma,\pi,\delta\}$ labels the bond symmetry, and the superscript
$(s)$ identifies the neighbor shell. Collecting the independent bond integrals
of Eq.~\eqref{eq:sk_hop} into a single vector $\{V_\lambda\}$ and defining the
design tensor
\begin{equation}
  M^{(s)}_{b\lambda\alpha\beta}
  = c_{\alpha\beta\mu(\lambda)}\!\left(\hat{\mathbf{d}}_b\right),
\label{eq:design_tensor}
\end{equation}
with $b$ a bond index, the Bloch Hamiltonian becomes
$H_{\alpha\beta}(\mathbf{k}) = \sum_\lambda V_\lambda
\sum_b M^{(s)}_{b\lambda\alpha\beta}\, e^{i\mathbf{k}\cdot\mathbf{d}_b}$.
Two consequences follow. First, the geometric factors are independent of the
parameters and need be evaluated only once for a given structure, yielding a
block-sparse representation that we exploit for computational efficiency.
Second, the Hamiltonian is \emph{linear} in the bond integrals, so the
derivatives required for gradient-based optimization are available in closed
form (Sec.~\ref{sec:SKTB:fit}).

\subsection{Environment dependence}
\label{sec:SKTB:env}

Conventional SK models are tied to the equilibrium geometry used to fit them and
degrade for surfaces, interfaces, and strained configurations in which the local
coordination differs from that reference. We therefore adopt the
environment-dependent tight-binding (EDTB) formalism of Tang \emph{et
al.}~\cite{Tang1996}, in which each bond integral is modulated by a screening
factor,
\begin{equation}
  \tilde{V}_\lambda^{(s)}(i,j)
  = V_\lambda^{(s)} \exp\!\left(-\gamma_\lambda S_{ij}\right),
\label{eq:screening}
\end{equation}
with the screening sum
\begin{equation}
  S_{ij} = \sum_{k \neq i,j} f_c(d_{ik})\, f_c(d_{jk}) .
\label{eq:screening_sum}
\end{equation}
Here $d_{ik}$ is the interatomic distance and $f_c$ is a cosine-tapered cutoff
function decreasing monotonically from unity at short range to zero beyond the
cutoff. $S_{ij}$ accumulates contributions from atoms $k$ that lie close to both
endpoints of the bond, so that a more crowded environment attenuates the hopping
exponentially, reproducing the reduction of covalent bonding strength under
increased coordination. The screening strengths $\gamma_\lambda$ may be
specified at three levels of granularity: a single global value, one per
angular-momentum pair $ll'$, or one per SK bond integral. On-site energies may
additionally be shifted by a coordination-dependent term $\eta_\alpha C_i$, with
$C_i = \sum_{k \neq i} f_c(d_{ik})$ the effective coordination number of atom
$i$.

For systems with continuously varying bond lengths, such as surfaces and
superlattices, the discrete-shell integrals are replaced by the
Goodwin-Skinner-Pettifor form~\cite{Goodwin1989},
\begin{equation}
  V_\lambda(r) = V_{0,\lambda}
  \left(\frac{r_0}{r}\right)^{n_\lambda}
  \exp\!\left\{ n_\lambda
    \left[ -\left(\frac{r}{r_c}\right)^{n_c}
           + \left(\frac{r_0}{r_c}\right)^{n_c} \right] \right\},
\label{eq:gsp}
\end{equation}
where $r_0$ is the reference bond length, $V_{0,\lambda}$ the corresponding bond
integral, $r_c$ the cutoff radius, and $n_\lambda$ and $n_c$ control the
power-law and exponential decay, respectively. Equation~\eqref{eq:gsp} vanishes
smoothly at $r_c$. For multispecies or multienvironment systems, each atom
carries an environment label; bonds between atoms of the same label draw on
independent parameter sets, while cross-environment bonds are assigned by a
geometric-mean mixing rule that preserves the sign structure of the SK
integrals.

\subsection{Parameter determination}
\label{sec:SKTB:fit}

All parameters, namely on-site energies, bond integrals, screening strengths, and
decay exponents, are obtained by minimizing the root-mean-square deviation
between the EDTB and PAO eigenvalues on a Brillouin-zone mesh. Minimization uses
the Levenberg-Marquardt algorithm with an analytic Jacobian supplied by the
Hellmann-Feynman theorem,
\begin{equation}
  \frac{\partial \varepsilon_{n\mathbf{k}}}{\partial p_j}
  = \left\langle n\mathbf{k} \middle|
    \frac{\partial H(\mathbf{k})}{\partial p_j}
    \middle| n\mathbf{k} \right\rangle ,
\label{eq:hf_jacobian}
\end{equation}
where $p_j$ denotes any fit parameter. Because $H(\mathbf{k})$ is linear in the
bond integrals, $\partial H/\partial V_\lambda$ is simply the corresponding
design tensor of Eq.~\eqref{eq:design_tensor}; the derivatives with respect to
the screening strengths and decay exponents follow analytically from
Eqs.~\eqref{eq:screening} and~\eqref{eq:gsp}. Finite-difference approximations
are thus avoided entirely, and convergence is typically reached within
$\mathcal{O}(10^2)$ iterations. Tikhonov regularization suppresses unphysical
oscillations in the fitted parameters, and multistart optimization guards
against convergence to local minima.

Transferability rests on multigeometry training. The residual is assembled from
band structures computed for $N_g$ configurations with distinct coordination
environments--for example, several isotropically strained bulk cells, or a slab
together with its bulk reference--and a single parameter set is fitted to all of
them simultaneously. This is necessary rather than merely advantageous: in a
uniform crystal every symmetry-equivalent bond shares the same $S_{ij}$, so the
exponential in Eq.~\eqref{eq:screening} reduces to a constant prefactor that is
degenerate with $V_\lambda^{(s)}$ and cannot be determined from a single
geometry. Training across geometries that differ in coordination lifts this
degeneracy and yields screening parameters with independent physical content.
The resulting models remain accurate for configurations outside the training
set - slabs, interfaces, superlattices, and large supercells - without further
DFT input.

\subsection{Application to twisted bilayer graphene}
\label{sec:SKTB:tbg}

\begin{figure}[!h]
    \centering
    \includegraphics[width=\linewidth]{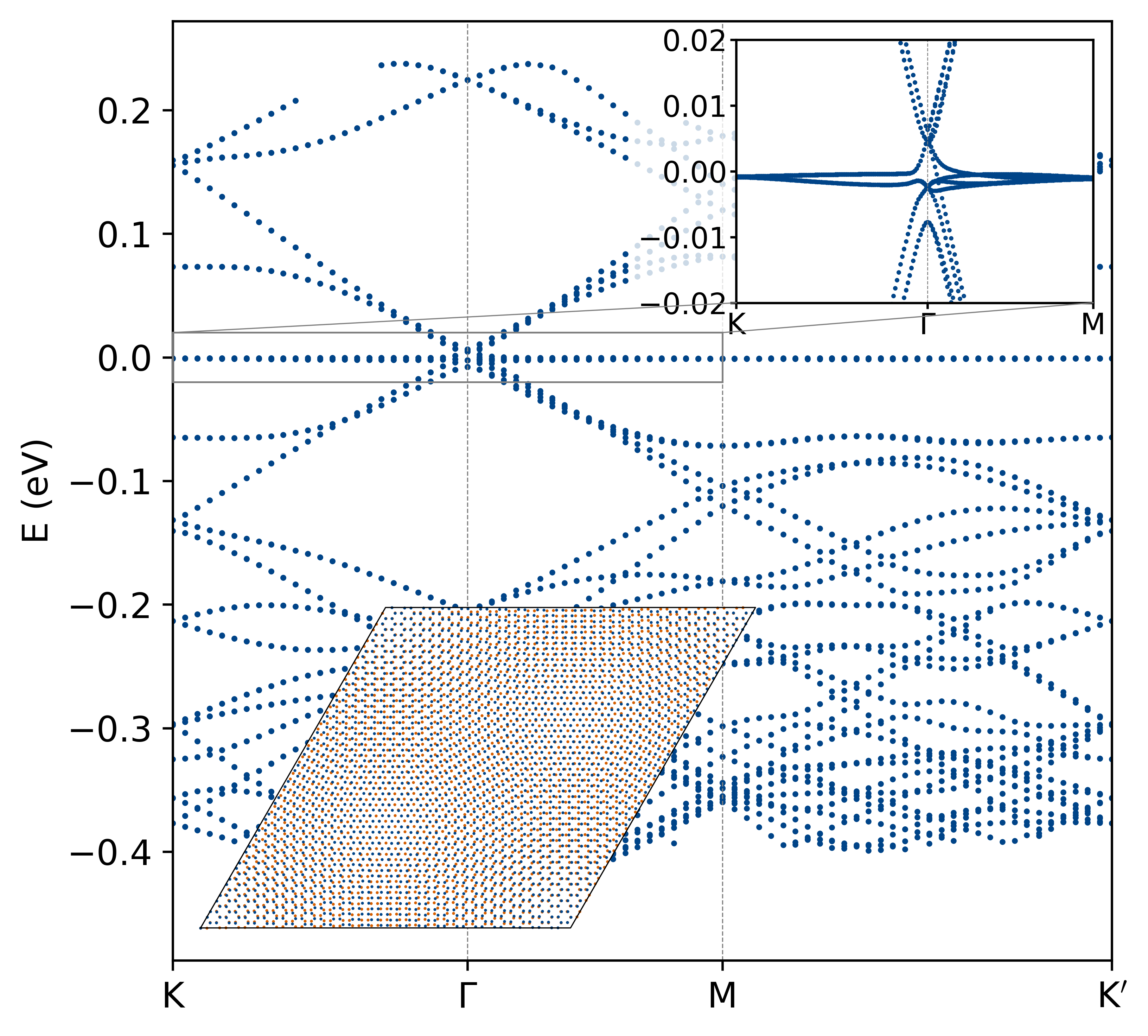}
    \caption{Sparse Lanczos band structure along $K$-$\Gamma$-$M$-$K'$ for the
    $(20,19)$ twisted bilayer graphene moir\'e superlattice. The lower-left inset
    shows the TBG geometry; the upper-right inset magnifies the low-energy region
    about $E = 0$~eV, resolving the nearly flat band manifold of
    $\sim$10-20~meV bandwidth. Adapted from Ref.~\cite{Marco2026EDTB} using the
    original data, under CC BY.}
    \label{fig:TBG}
\end{figure}

The $(20,19)$ twisted bilayer graphene (TBG) moir\'e superlattice, containing
4324 atoms (Fig.~\ref{fig:TBG}), illustrates the method at scale. Mapping the
local atomic relaxations onto SK integrals through the PAO eigenvalue reference
yields a transferable Hamiltonian for a system well beyond the reach of direct
DFT calculation.

At this size, dense eigensolvers are impractical, and we instead exploit the
intrinsic sparsity of the EDTB Hamiltonian. Matrix elements are constructed on
the fly by bond enumeration, using the fitted distance-dependent and screening
parameters without reparametrization. At each $\mathbf{k}$ point along the
moir\'e Brillouin-zone path we compute the 40 eigenvalues nearest the
shift-invert target $\sigma = 0$~eV, corresponding to the Dirac point, by the
implicitly restarted Lanczos algorithm~\cite{Lanczos1950,Sorensen1992}. For the
largest calculation reported here this reduces the memory requirement from
approximately $11$~GB to $250$~MB, placing systems of $10^5$ orbitals within
reach of a single workstation~\cite{Marco2026EDTB}.

\section{Transport Properties}
\label{sec:transport}

\subsection{Semiclassical Boltzmann transport}
\label{sec:transport:boltzmann}

\paoflow{} implements the semiclassical Boltzmann transport formalism for calculating the electrical conductivity, Seebeck coefficient, electronic contribution to thermal conductivity, as well as the Hall and Nernst coefficients. The implementation is based on the linearized Boltzmann transport equation within the relaxation time approximation (RTA), in which the transport coefficients are conveniently expressed in terms of the transport distribution function (TDF)~\cite{Ziman1960,Madsen2006}: 

\begin{equation}
\label{eq:transport:TDF}
\Xi^{ij}(\varepsilon)=
\frac{1}{V_{\mathrm{BZ}}}
\sum_n
\int_{\mathrm{BZ}} d^3k\,
v_{n\mathbf{k}}^{\,i}
v_{n\mathbf{k}}^{\,j}
\,\tau_{n\mathbf{k}}\,
\delta(\varepsilon-\varepsilon_{n\mathbf{k}})
\end{equation}

%

\noindent where \(v_{n\mathbf{k}}^{\,i} = \hbar^{-1} \partial\varepsilon_{n\mathbf{k}}/\partial k_i\) is the $i$-th component of the group velocity, and $\tau_{n\mathbf{k}}$ is the relaxation time, which may be taken as a constant or specified as a function of energy. To improve the convergence of the TDF, we implement an adaptive broadening of the $\delta(\varepsilon - \varepsilon_{n\mathbf{k}})$ term which permits the calculation of transport properties at low temperatures ($T$), without requiring excessively dense $k$-meshes. The examples of convergence improvement thanks to the adaptive broadening are shown in Fig. \ref{fig:boltzmann}.

\subsubsection{Relaxation time models}
\label{sec:transport:relaxation_model}

\begin{table*}[t]
  \centering
  \small
  \begin{tabular}{@{}l@{\hspace{3em}}l@{\hspace{3em}}l@{}}
    \toprule
    Mechanism & $E$, $T$ dependence & Parameters \\
    \midrule
    Acoustic phonon &
      $T^{-1}E^{-1/2}$ &
      $\rho$, $v$, $m^{*}$, $D_{\mathrm{ac}}$ \\[2pt]
    Optical phonon &
      $T^{1/2}\left[N_{\mathrm{op}}\sqrt{x+x_{o}}
        + (N_{\mathrm{op}}\!+\!1)\Theta(x-x_{o})\sqrt{x-x_{o}}\right]^{-1}$ &
      $\rho$, $m^{*}$, $D_{\mathrm{op}}$, $\hbar\omega_{\mathrm{op}}$ \\[2pt]
    Polar optical &
      $E^{3/2}$, summed over LO branches &
      $\epsilon_{0}$, $\epsilon_{\infty}$, $m^{*}$, $\hbar\omega_{\mathrm{LO}}$ \\[2pt]
    Ionized impurity &
      $E^{3/2}\left[\ln(1+x^{-1})-(1+x)^{-1}\right]^{-1}$ &
      $\varepsilon$, $m^{*}$, $n_{I}$, $Z_{I}$ \\[2pt]
    Piezoelectric &
      $T^{-1}E^{1/2}$ &
      $p$, $\varepsilon$, $\rho$, $v$, $m^{*}$ \\
    \bottomrule
  \end{tabular}
  \caption{Relaxation time models implemented in \paoflow{}~\cite{Jayaraj2022}. The reduced variables are $x = E/k_{B}T$ and $x_{o} = \hbar\omega_{\mathrm{op}}/k_{B}T$, $N_{\mathrm{op}} = [\exp(x_{o})-1]^{-1}$ is the Bose-Einstein population of the optical mode, and $\varepsilon = \epsilon_{0} + \epsilon_{\infty}$. $D_{\mathrm{oc}}$ and $D_{\mathrm{op}}$ are the acoustic and optical deformation potentials, respectively, and $\omega_{\mathrm{oc}}$ and $\omega_{\mathrm{LO}}$ denote the optical phonon and longitudinal-optical phonon frequencies. $\epsilon_{0}$ and $\epsilon_{\infty}$ are the static and high-frequency dielectric constants, $n_{I}$ is the impurity concentration, $Z_{I}$ is the impurity charge state and $p$ is the piezoelectric constant. 
    }
  \label{tab:transport:rta_models}
\end{table*}

The constant relaxation time approximation (CRTA), in which $\tau$ is treated as a single adjustable parameter, is commonly used because it can be factored out of the transport integrals based on Eq.~(\ref{eq:transport:TDF}), leaving the transport coefficients parameter-free. However, it neglects the energy and temperature dependence of scattering, which is essential to reproduce experimental trends even in simple semiconductors. \paoflow{} therefore implements a hierarchy of analytical relaxation-time models built on semiclassical scattering theory~\cite{jacoboni2010theory}, following the strategy of Ref.~\cite{Farris2018theory} and described in detail in Ref.~\cite{Jayaraj2022}.  These models capture the dominant intrinsic and extrinsic scattering mechanisms at negligible computational cost, making them suitable for high-throughput calculations, where calculations of electron-phonon self-energies are prohibitively expensive.

Each scattering mechanism contributes a relaxation time $\tau(E,T)$ that depends on the band energy $E$, and on the temperature $T$. Acoustic-phonon scattering, treated in the elastic deformation-potential limit, is representative of the family:
\begin{equation}
  \tau_{\mathrm{ac}}(E,T)
  = \frac{2\pi\hbar^{4}\rho v^{2}}
         {(2m^{*})^{3/2}\,k_{B}T\,D_{\mathrm{ac}}^{2}\,\sqrt{E}} 
  \label{eq:transport:tau_ac}
\end{equation}
where $\rho$ is the mass density, $v$ the acoustic velocity, $m^{*}$ the effective-mass ratio and $D_{\mathrm{ac}}$ the acoustic deformation potential. The remaining mechanisms, including nonpolar optical, polar optical, ionized
impurity and piezoelectric scattering share the same structure and are collected in Table~\ref{tab:transport:rta_models} together with their characteristic energy and temperature dependence and the required input parameters. Explicit expressions are given in Ref.~\cite{Jayaraj2022}.

The models are evaluated on the interpolated PAO band structure and enter the transport integrals in a band- and $\mathbf{k}$-resolved form, $\tau_{n\mathbf{k}} = \tau\!\left(\epsilon_{n}(\mathbf{k}),T\right)$, so that the
full anisotropy and non-parabolicity of $\epsilon_{n}(\mathbf{k})$ are retained in Eq.~(\ref{eq:transport:TDF}) even though the analytical $\tau(E,T)$ expressions are themselves derived within a parabolic-band, long-wavelength
phonon approximation.  This is the standard assumption for semiconductors, where carriers occupy narrow energy windows around the band extrema, but it should be applied with care to systems in which short-wavelength phonon scattering or
strongly non-parabolic dispersion dominates the transport.

The individual channels are combined according to Matthiessen's rule, $\tau_{\mathrm{total}}^{-1} = \sum_{i}\tau_{i}^{-1}$.  The material-specific parameters ($\rho$, $v$, $m^{*}$, $D_{\mathrm{ac}}$, $D_{\mathrm{op}}$, $\hbar\omega_{\mathrm{op}}$, $\epsilon_{0}$, $\epsilon_{\infty}$, $n_{I}$, $Z_{I}$, $p$) may be determined from experiment or from independent first-principles calculations.  Alternatively, \paoflow{} provides a self-consistent fitting mode in which these parameters are held fixed and a set of dimensionless, temperature-dependent correction functions $a_{i}(T)$ is introduced through a modified Matthiessen rule:

\begin{equation}
\begin{split}
\frac{1}{\tau_{\mathrm{total}}(E,T)}
&=
\frac{a_{\mathrm{imp}}(T)}{\tau_{\mathrm{imp}}(E,T)}
+\frac{a_{\mathrm{ac}}(T)}{\tau_{\mathrm{ac}}(E,T)}
+\frac{a_{\mathrm{op}}(T)}{\tau_{\mathrm{op}}(E,T)}
\\
&\quad
+\frac{a_{\mathrm{pop}}(T)}{\tau_{\mathrm{pop}}(E,T)}
+\frac{a_{\mathrm{pac}}(T)}{\tau_{\mathrm{pac}}(E,T)} .
\end{split}
\label{eq:transport:modified_matthiessen}
\end{equation}

The correction functions $a_i(T)$ are optimized by sequential least-squares programming~\cite{nocedal2006sequential} so that the calculated electrical conductivity reproduces the measured $\sigma(T)$ of a given sample.  Because each scattering channel carries its own correcting function, the fit yields more than agreement with the data: the magnitude of $a_{i}(T)$ quantifies how far each base model departs from the behaviour of that particular sample.  Correcting functions close to unity indicate that the analytical models already provide a faithful description of the scattering, whereas a large $a_{i}(T)$ identifies a channel whose strength the base model underestimates~\cite{Jayaraj2022}.  This approach provides a flexible route to sample-specific scattering information, including extrinsic contributions from dopants, grain boundaries or disorder, without recourse to full electron-phonon calculations.

\subsubsection{Electrical and thermoelectric transport}
\label{sec:transport:sigma_seebeck}
Within the semiclassical Boltzmann formalism, the charge current density may be expressed as a linear response to an external electric field, magnetic field, and temperature gradient according to:

\begin{equation}
j^{i}
=
\sigma^{ij}E_{j}
+
\sigma_H^{ijk}E_{j}B_{k}
+
\nu^{ij}\nabla_{j}T
+
\nu_N^{ijk}\nabla_{j}TB_{k} + ...
\end{equation}

\noindent where $\sigma^{ij}$ and $\sigma_H^{ijk}$ denote the electrical conductivity and Hall conductivity tensors, respectively, while $\nu^{ij}$ and $\nu_N^{ijk}$ are the corresponding thermoelectric and Nernst conductivity tensors. These response functions are conveniently expressed in terms of the generating transport tensors $\mathcal{L}^{ij}_{\alpha}$, with $\alpha=0,1,2$, defined as~\cite{Mecholsky2014}:

\begin{equation}
  \mathcal{L}^{ij}_{\alpha} =
   \frac{1}{V_{\mathrm{BZ}}}
\sum_n
\int_{\mathrm{BZ}} d^3k\,
v_{n\mathbf{k}}^{\,i}
v_{n\mathbf{k}}^{\,j}
\,\tau_{n\mathbf{k}}
    (\varepsilon_{n\mathbf{k}}-\mu)^\alpha\, 
    \left(-\frac{\partial f}{\partial \varepsilon}\right)
\end{equation}

\noindent where $f(\varepsilon;\mu,T)$ is the Fermi–Dirac distribution and $\mu$ is the chemical potential. If an adaptive broadening is used, first the TDF is computed and then the generating tensors are obtained integrating over $\varepsilon$~\cite{Madsen2006}:

\begin{equation}
\mathcal{L}^{ij}_{\alpha}
=
\int d\varepsilon\,
(\varepsilon-\mu)^{\alpha}
\,
\Xi^{ij}(\varepsilon)
\left(
-\frac{\partial f}{\partial\varepsilon}
\right),
\end{equation}


The electrical conductivity and thermoelectric conductivity tensors are obtained directly from the generating transport tensors as~\cite{Bies2002}

\begin{align}
\sigma^{ij}(T,\mu)
&=
e^{2}\mathcal{L}^{ij}_{0},
\\[4pt]
\nu^{ij}(T,\mu)
&=
\frac{e}{T}\mathcal{L}^{ij}_{1}.
\end{align}

The experimentally measured Seebeck coefficient and electronic contribution to the thermal conductivity at zero electrical current are then given by~\cite{Bies2002}:

\begin{align}
S^{ij}(T,\mu)
&=
-\left(
{\sigma}^{-1}
{\nu}
\right)^{ij}
=
-\frac{1}{eT}
\left(
\mathcal{L}_{0}^{-1}
\mathcal{L}_{1}
\right)^{ij},
\\[4pt]
\kappa^{ij}_{e}(T,\mu)
&=
\frac{1}{T}
\left(
\mathcal{L}_{2}
-
\mathcal{L}_{1}
\mathcal{L}_{0}^{-1}
\mathcal{L}_{1}
\right)^{ij}.
\end{align}



\subsubsection{Ordinary Hall effect}
\label{sec:transport:ordinary_hall}

Within the relaxation time approximation and assuming an isotropic relaxation time, the Hall conductivity is obtained in the low field limit from~\cite{Madsen2006}:

\begin{align}
\label{eq:transport:THDF}
\Xi_H^{ijk}(\varepsilon)
=
\frac{1}{V_{\mathrm{BZ}}}
\sum_n
\int_{\mathrm{BZ}} d^3k\,
\tau_{n\mathbf{k}}^{2}\,
\delta(\varepsilon-\varepsilon_{n\mathbf{k}})
\\
\times
\sum_{lm}
\epsilon_{klm}\,
v^{i}_{n\mathbf{k}}\,
v^{m}_{n\mathbf{k}}\,
M^{-1}_{jl,n\mathbf{k}}
\end{align}

\noindent where $\epsilon_{klm}$ is the Levi-Civita tensor and

\begin{equation}
M^{-1}_{ij,n\mathbf{k}}
=
\frac{1}{\hbar^{2}}
\frac{\partial^{2}\varepsilon_{n\mathbf{k}}}
{\partial k_i\,\partial k_j}
\end{equation}

\noindent is the inverse effective mass tensor. The Hall conductivity tensor $\sigma_H$ is then obtained by integrating over energy:

\begin{equation}
\sigma_H^{ijk}(T,\mu)
=
e^{3}
\int d\varepsilon\,
\Xi_H^{ijk}(\varepsilon)
\left(
-\frac{\partial f}{\partial\varepsilon}
\right).
\end{equation}



In the low field limit, the Hall voltage follows a linear dependence with magnetic field

\begin{equation}
E_i
=
R_H^{ijk}
J_j
B_k,
\end{equation}

\noindent with a slope equal to the Hall coefficient $R_H$~\cite{Madsen2006}. $R_H$ is a constant independent of magnetic field and relaxation time, and it can be obtained by inverting the magneto-conductivity tensor: 

\begin{equation}
  \rho(B) = (\sigma(B))^{-1} = (\sigma_0 +B\sigma_{H} + \mathcal{O}(B^2) )^{-1}
\end{equation}

\noindent where $ \sigma_0 = \sigma(B=0)$. The $R_H$ coefficient corresponds to the linear term in expansion series around $B=0$:
\begin{equation}
  R_H = -\sigma_0^{-1} \sigma_{H} \sigma_0^{-1}
\end{equation}

\subsubsection{Nernst effect}
\label{sec:transport:nernst}

Within the isotropic RTA, the Nernst thermoelectric conductivity tensor $\nu_N^{ijk}$ is obtained analogously to the Hall conductivity tensor as:

\begin{equation}
\nu_N^{ijk}(T,\mu)
=
\frac{e^2}{T}
\int d\varepsilon\,
(\varepsilon-\mu)
\Xi_H^{ijk}(\varepsilon)
\left(
-\frac{\partial f}{\partial\varepsilon}
\right)
\end{equation}

The field independent Nernst coefficient is obtained by linearizing the thermopower tensor $\theta$ with respect to the applied magnetic field~\cite{Pikulin2011}: 

\begin{equation}
\theta(B)
=
\rho(B)\nu(B)
=
\rho(B)
\left(
\nu
+
B\nu_N
+
\mathcal{O}(B^2)
\right),
\end{equation}

\noindent where $\rho(B)=\sigma(B)^{-1}$ and $\nu=-\sigma S$ is the zero-field thermoelectric conductivity tensor. The field-independent Nernst coefficient is then given by:

\begin{equation}
\mathcal{N}
=
\sigma_0^{-1}
\nu_N
+R_H \nu.
\end{equation}

In the low field limit, the Nernst coefficient, which is the analogue of the Seebeck coefficient in the presence of a magnetic field, relates the electric field to the temperature gradient and magnetic field according to:

\begin{equation}
E_i
=
\mathcal{N}^{ijk}
\nabla_jT
B_k.
\end{equation}










\subsubsection{Rashba-Edelstein effect}
\label{sec:transport:rashba}

In gyrotropic materials, an applied electric field induces not only a charge current but also a net spin accumulation, which is known as the Rashba-Edelstein effect~\cite{Ganichev2002, Longrange2022}. In analogy to Eq.~(\ref{eq:transport:TDF}), the spin accumulation can be expressed in terms of: 

\begin{equation}
\Xi_S^{ij}(\varepsilon)
=
\frac{e}{V_{\mathrm{BZ}}}
\sum_n
\int_{\mathrm{BZ}} d^3k\,
\tau_{n\mathbf{k}}\,
\langle S^{i}\rangle_{n\mathbf{k}}\,
v^{j}_{n\mathbf{k}}\,
\delta(\varepsilon-\varepsilon_{n\mathbf{k}})
\end{equation}

\noindent where $\langle S^{i}\rangle_{n\mathbf{k}}$ denotes the expectation value of the spin operator. The nonequilibrium spin accumulation per unit volume is then obtained as:

\begin{equation}
\delta s^{i}
=
E_j
\int d\varepsilon\,
\Xi_S^{ij}(\varepsilon)
\left(
-\frac{\partial f}{\partial\varepsilon}
\right)
\end{equation}

\noindent which has the same structure as the transport coefficients. As in the case of electrical transport, only electronic states in the vicinity of the Fermi level contribute through the derivative of the Fermi-Dirac distribution.

To eliminate the explicit dependence on the relaxation time, the current-induced spin accumulation is expressed in terms of the REE susceptibility tensor as~\cite{Analogs2023, Tenzin2023}:

\begin{equation}
\delta s^{i}
=
\chi^{ij}
j^{j},
\end{equation}

\noindent where $j^{j}$ is the applied charge current, and $\chi^{ij}$ is:

\begin{equation}
\chi^{ij}
=
-
\frac{
\displaystyle
\sum_n
\int_{\mathrm{BZ}} d^3k\,
\langle S^{i}\rangle_{n\mathbf{k}}\,
v^{j}_{n\mathbf{k}}
\left(
-\frac{\partial f}{\partial\varepsilon}
\right)
}{
\displaystyle
e
\sum_n
\int_{\mathrm{BZ}} d^3k\,
\left(
v^{j}_{n\mathbf{k}}
\right)^2
\left(
-\frac{\partial f}{\partial\varepsilon}
\right)
}
\label{eq:transport:ree}
\end{equation}

\begin{figure}[!ht]
    \centering
    \includegraphics[width=\linewidth]{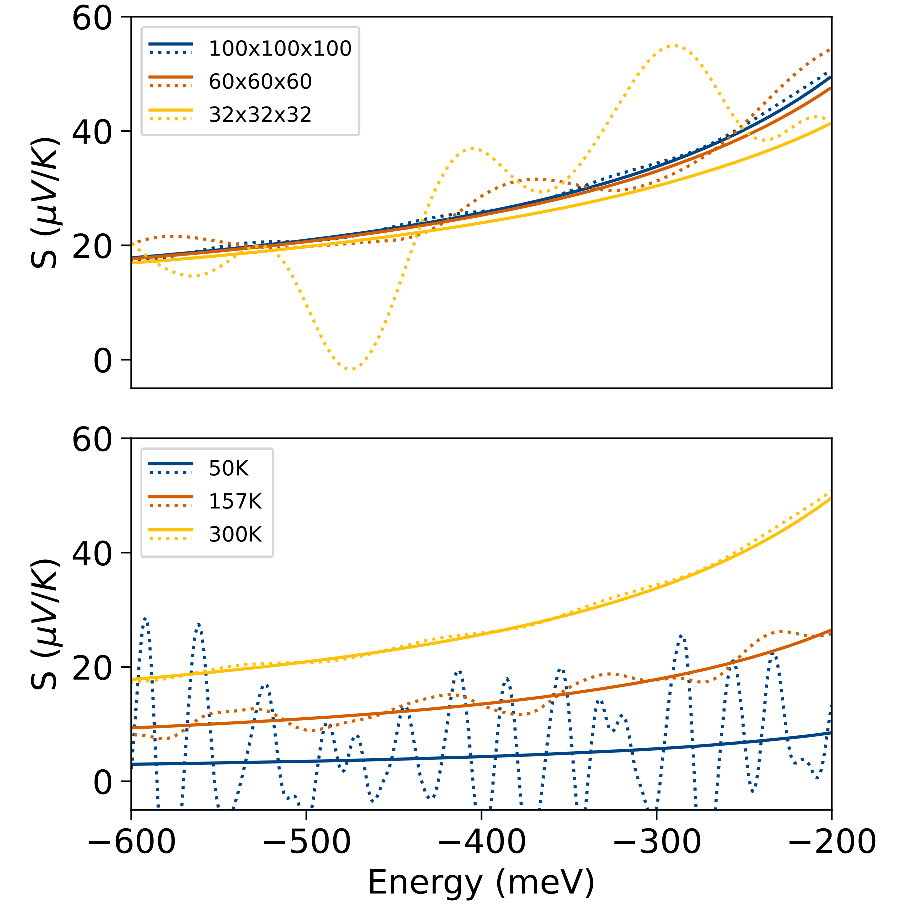}
    \caption{Effect of adaptive broadening for the calculation of the Seebeck effect in silicon. Top: Seebeck effect at 300K for different mesh densities. Bottom: Seebeck effect at different temperatures for a fine mesh of 100$^3$. Solid (dotted) lines correspond to calculations with (without) adaptive broadening.}
    \label{fig:boltzmann}
\end{figure}


\subsection{Kubo formalism}
\label{sec:transport:kubo}

In addition to the semiclassical transport, \paoflow{} implements the Kubo linear response formalism for calculating the electrical conductivity, anomalous Hall conductivity (AHC)~\cite{Jungwirth2002, Yao2004}, spin Hall conductivity (SHC)~\cite{Sinova2004, Guo2008}, and Rashba-Edelstein effects~\cite{Manchon2015, GonzalezHernandez2024}. While the electrical conductivity can also be evaluated within the semiclassical formalism, the Kubo implementation additionally provides access to the intrinsic AHC and SHC, which are not commonly treated within the semiclassical approach~\cite{Sinitsyn2008}. In the constant relaxation time approximation, the response of an observable $\delta\mathbf{A}$ induced by an external electric field $\mathbf{E}$ is expressed as:

\begin{equation}
\delta A_i = (\chi ^I_{ij} + \chi^{II}_{ij}) E_j
\end{equation}

\noindent with the two contributions $\chi_I$ and $\chi_{II}$ defined as:

\small
\begin{eqnarray}
\chi_{i j}^I=-\frac{e \hbar}{\pi} \sum_{\mathbf{k}, n,m} \frac{\Gamma ^2\operatorname{Re}\left[\left\langle\psi_{\mathbf{k} n}\right| \hat{A}_i\left|\psi_{\mathbf{k} m}\right\rangle\left\langle\psi_{\mathbf{k} m}\right| \hat{v}_j\left|\psi_{\mathbf{k} n}\right\rangle\right]}{\left(\left(E_F-E_{\mathbf{k} n}\right)^2+\Gamma^2\right)\left(\left(E_F-E_{\mathbf{k} m}\right)^2+\Gamma^2\right)}\nonumber\\
\end{eqnarray}

\begin{equation}
\chi_{i j}^{I I} \approx-2 e \hbar \sum^{\substack{\textit{\tiny{n~occ.}} \\ \textit{\tiny{m~unocc.}}}}_{\mathbf{k},n\neq m} \frac{\operatorname{Im}\left[\left\langle\psi_{k n}\right| \hat{A}_i\left|\psi_{k m}\right\rangle\left\langle\psi_{k m}\right| \hat{v}_j\left|\psi_{k n}\right\rangle\right]}{\left(E_{k n}-E_{k m}\right)^2}
\label{final_chi_ii}
\end{equation}

\normalsize

\noindent In the above equations, $e$ is the elementary (positive) charge, $\mathbf{k}$ is the Bloch wave vector, $n$ and $m$ are the band indices, $E_{\mathbf{k}n}$ is the band energy, $E_F$ is the Fermi energy, $\hat{v}$ is the velocity operator, and $\Gamma$ is the disorder broadening related to the relaxation time $\tau$ through $\Gamma=\hbar/(2\tau)$. The expression for $\chi^{II}$ corresponds to the low-disorder limit of the Kubo formalism, obtained by neglecting terms of order $\Gamma^2$, and represents the intrinsic contribution implemented in \paoflow{}.

The response tensor is evaluated for different choices of the operator $\hat{A}$. Choosing the velocity operator, $\hat{A}=\hat{v}$, yields the electrical conductivity through $\chi^I$ and, in magnetic materials, the intrinsic anomalous Hall conductivity through $\chi^{II}$. Choosing the spin operator, $\hat{A}=\hat{s}$, yields the Rashba-Edelstein effect, for which the contributions $\chi^I$ and $\chi^{II}$ correspond to the $\mathcal{T}$-even and $\mathcal{T}$-odd components, respectively~\cite{Tenzin2025}. Replacing the spin operator by the spin current operator constructed as:

\begin{equation} \hat{J}^i_{l} = \frac{1}{2}\left\{\hat{s}_i, \hat{v}_l\right\}\label{eq:scurrent}\end{equation}

\noindent yields the spin Hall conductivity. In this case, $\chi^{II}$ corresponds to the conventional $\mathcal{T}$-even intrinsic SHC present in nonmagnetic materials, whereas $\chi^I$ describes the magnetic ($\mathcal{T}$-odd) spin Hall effect that can emerge in materials without time-reversal symmetry~\cite{Tenzin2025, Ryosuke2026}. 

In addition, \paoflow{} supports calculations in which the spin operators are replaced by orbital angular momentum operators within the atom-centered approximation, following the previous studies~\cite{Dongwook2018, Marcio2021}.

\begin{figure}
    \centering
    \includegraphics[width=\linewidth]{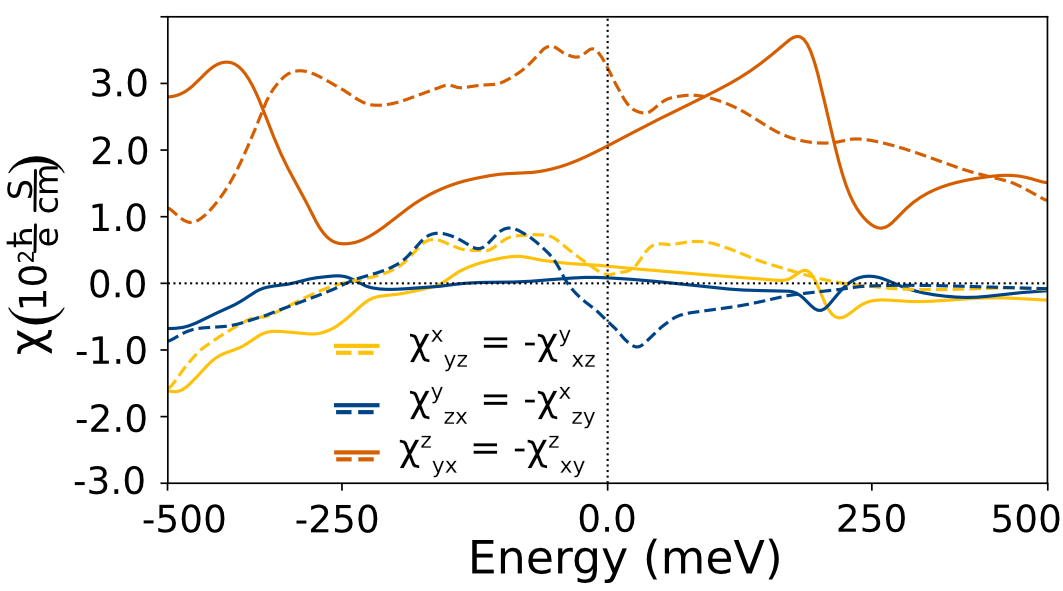}
    \caption{Spin Hall conductivity in nonmagnetic (dashed lines) and altermagnetic (solid lines) NiTa$_3$S$_6$ calculated as a function of chemical potential. In the antiferromagnetic phase, the Néel vector is oriented along the out-of-plane direction. Only $\mathcal{T}$-even components are present, as $\mathcal{T}$-odd components are forbidden by crystal symmetries in both phases. Adapted from Ref.~\cite{Tenzin2025} using the original data, under CC BY.}
    \label{fig:kuboshe}
\end{figure}

\subsection{Projected response functions}
\label{sec:topo:sitebands}
Each PAO basis function carries an
unambiguous atomic and orbital label. Therefore, any partition of the basis into disjoint sets induces a corresponding decomposition of the electronic Hilbert space. For a set $\tau$, where $\tau$ may represent a group of orbitals from an atomic layer or a single atomic site, we define the corresponding projector as
\begin{equation}
  P_{\tau} \;=\; \sum_{i \in \tau} \ket{i}\bra{i},
  \qquad
  \sum_{\tau} P_{\tau} = \mathbf{1},
  \label{eq:projector_s}
\end{equation}
which in this basis is a diagonal matrix carrying unit entries on the selected orbitals and zeros elsewhere. The decomposition is applied to the operator rather than to the electronic states. For example, the portion of the spin
current $\hat{J}^{i}_{l}$ (Eq. \eqref{eq:scurrent}) carried by a set $\tau$ is defined by the symmetrized projection
\begin{equation}
  \hat{J}^{i}_{l}(\tau)
  \;=\; \tfrac{1}{2}\,\anticomm{P_{\tau}}{\hat{J}^{i}_{l}}.
  \label{eq:projected_current_s}
\end{equation}
The symmetrized form is necessary because $P_{\tau}$ and $\hat{J}^{i}_{j}$ do not commute, which is generally true for all operators: only the anti-commutator remains Hermitian and yields a real response. In block form, each set retains its own intra-set block in full and exactly one half of each inter-set block, so that
\begin{equation}
  \sum_{\tau} \hat{J}^{i}_{j}(\tau) = \hat{J}^{i}_{j},
  \label{eq:sumrule_s}
\end{equation}
and the resolved contributions are additive, free of double counting, and with no residual term remaining.

Because $P_{\tau}$ acts on the operator, and not on the Hamiltonian, the
eigenvalues and eigenvectors entering the linear response expression remain
those of the complete system. Hybridization between the set of orbitals is, therefore, fully retained. The sum rule (Eq.~\eqref{eq:sumrule_s}) is inherited by the linear response, i.e. the projected spin Hall conductivity $\sum_{\tau}\sigma(\tau) = \sigma$. This construction is independent
of the operator being resolved: for instance, replacing the spin operator in $\hat{J}^{i}_{j}$ with the orbital angular momentum operator, or replacing the current operator in $\hat{J}^{i}_{j}$ with a density operator, yields correspondingly set-resolved orbital Hall conductivity or spin density. The unprojected result is recovered exactly by choosing $P_{\tau} = \mathbf{1}$. 

\begin{figure}[h!]
    \centering
    \includegraphics[width=\linewidth]{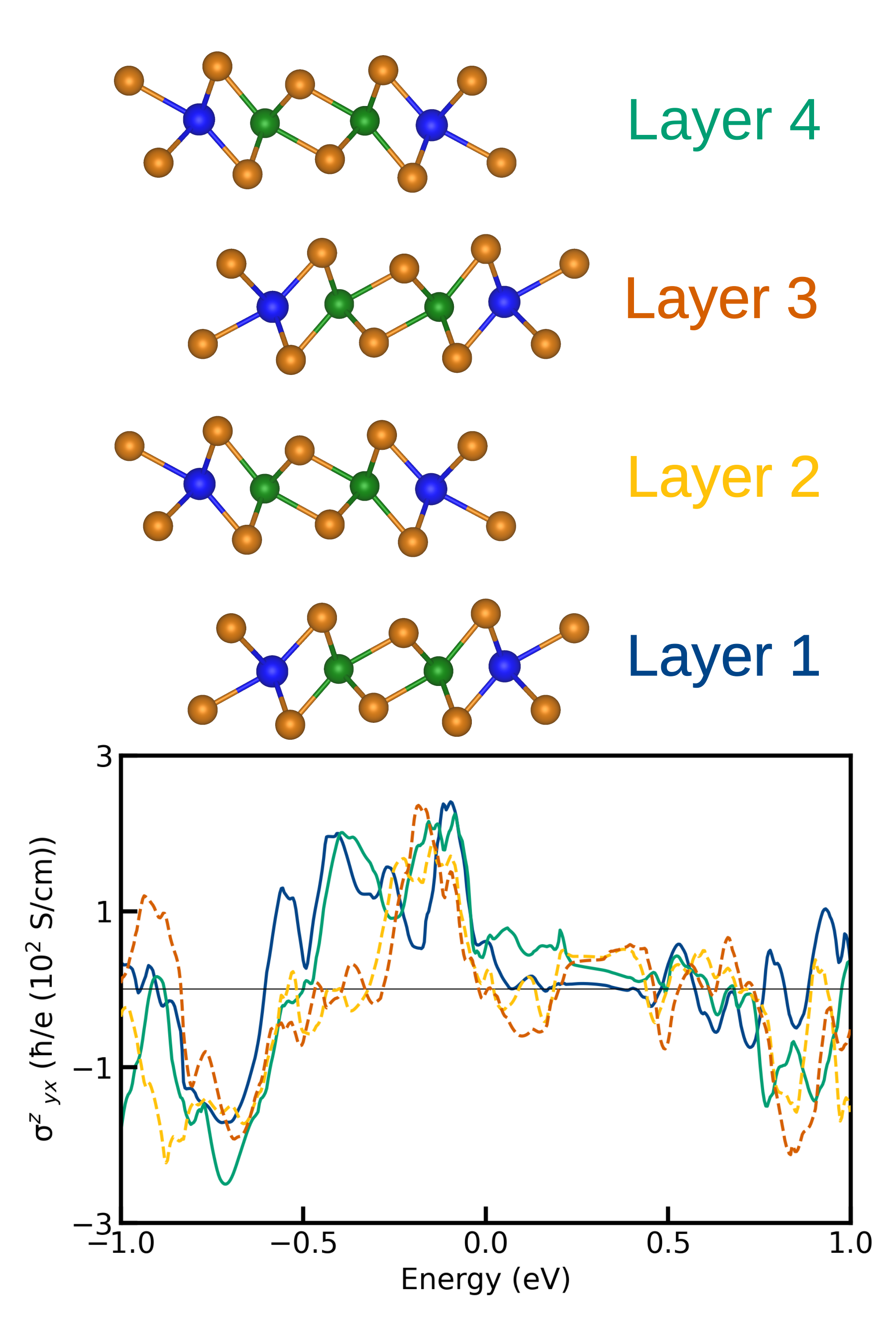}
    \caption{Individual layer contributions to spin Hall conductivity ($\sigma^z_{yx}$) of 4-layer NbIrTe$_4$ thin film. The dashed lines show the contributions of the inner layers, revealing their different behavior compared with the surface layers (solid lines).}
    \label{fig:NIT_layerProj}
\end{figure}

An example is shown in Fig.~\ref{fig:NIT_layerProj}, where the layer-projected spin Hall conductivity is calculated for a slab of NbIrTe$_4$ consisting of four van der Waals layers.

\subsection{Landauer-Büttiker quantum transport}
\label{sec:transport:landauer}

\paoflow{} implements coherent quantum transport in the Landauer-B\"uttiker formalism~\cite{Landauer1957,Buttiker1986,Datta1995}, complementing the semiclassical Boltzmann module of Sec.~\ref{sec:transport:boltzmann} with a fully quantum-mechanical description suited to nanoscale conductors, heterostructures and interfaces. The implementation follows the Green's-function scheme of Ref.~\cite{Nardelli1999}, previously realized on a maximally-localized Wannier basis~\cite{Calzolari2004}.  The method applies to any Hamiltonian that is short-ranged in a localized-orbital representation; the PAO Hamiltonian satisfies this by construction, being minimal and atom-centered, so that the same $H(\mathbf{R})$ used for band interpolation feeds directly into the transport calculation.

\begin{figure*}[t]
    \centering
    \includegraphics[width=0.9\textwidth]{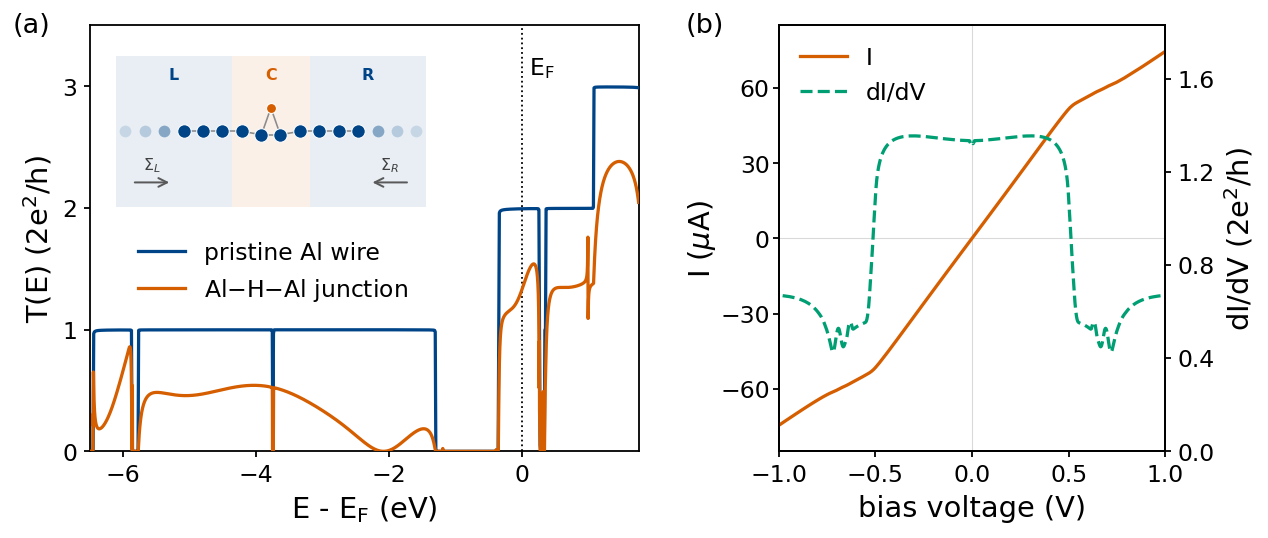}
    \caption{Transport through an Al monoatomic wire with a single H impurity. (a) Transmission of the pristine wire (blue quantized plateaus) versus the junction (orange). The inset shows the L/C/R partition of the device where the blue and orange atoms represent Al and H respectively. The leads are semi-infinite and enter through $\Sigma_{L,R}$. (b) $I(V)$ of the junction (left axis) and $dI/dV$ (right axis).}
    \label{fig:landauer}
\end{figure*}

The system is partitioned into a finite conductor $C$ connected to semi-infinite left ($L$) and right ($R$) leads, as shown in the inset of Fig.~\ref{fig:landauer} (a) for an Al monoatomic wire hosting a single H impurity. Each lead is treated as a stack of principal layers~\cite{LeeJoannopoulos1981}, defined so that only adjacent layers interact; the semi-infinite region is then completely specified by one on-site and one hopping block and can be folded exactly into a finite, energy-dependent self-energy.  For each Bloch vector $\mathbf{k}_{\parallel}$ transverse to the transport axis, the retarded Green's function of the conductor reads
\begin{equation}
  G^{r}(\mathbf{k}_{\parallel},E) =
    \bigl[(E+i\eta)\,\mathbb{I} - H_{C}(\mathbf{k}_{\parallel})
    - \Sigma^{r}_{L} - \Sigma^{r}_{R}\bigr]^{-1},
  \label{eq:transport:Gr}
\end{equation}
where $H_{C}$ is the conductor Hamiltonian, $\eta$ is a small positive value that smooths the spectral features, and the identity replaces the overlap matrix since the PAO projection yields an orthogonal representation. The lead self-energies follow from the surface Green's functions $g^{r}_{L/R}$ of the isolated semi-infinite leads,
\begin{equation}
  \Sigma^{r}_{L} = h^{\dagger}_{LC}\,g^{r}_{L}\,h_{LC},
  \qquad
  \Sigma^{r}_{R} = h_{CR}\,g^{r}_{R}\,h^{\dagger}_{CR},
  \label{eq:transport:sigma}
\end{equation}
with $h_{LC}$ and $h_{CR}$ the coupling matrices between the conductor and the adjacent lead layers. These terms act as effective, non-Hermitian Hamiltonians that encode the openness of the system.  The surface Green's functions are evaluated through the transfer matrices of the semi-infinite stacks, obtained by the iterative recursion of L\'opez Sancho \textit{et al.}~\cite{LopezSancho1984,Sancho1985}.

The transmission function is given by the Fisher-Lee relation~\cite{FisherLee1981},
\begin{equation}
\begin{split}
T(\mathbf{k}_{\parallel},E)
    &= \mathrm{Tr}\bigl[\Gamma_{L}\,G^{r}\,\Gamma_{R}\,G^{a}\bigr], \\
\Gamma_{L/R}
    &= i\bigl[\Sigma^{r}_{L/R}
      - (\Sigma^{r}_{L/R})^{\dagger}\bigr].
\end{split}
\label{eq:transport:fisherlee}
\end{equation}
where $G^{a}=(G^{r})^{\dagger}$ is the advanced Green's function and the coupling matrices $\Gamma_{L/R}$ describe the broadening of the conductor levels induced by the leads. Because the principal layer construction reduces the three-dimensional problem to a set of independent linear chains, one for each $\mathbf{k}_{\parallel}$, the total transmission is recovered by summation over the transverse Brillouin zone,
\begin{equation}
  T(E) = \sum_{\mathbf{k}_{\parallel}}
         w_{\mathbf{k}_{\parallel}}\,T(\mathbf{k}_{\parallel},E),
  \label{eq:transport:kpar}
\end{equation}
with $w_{\mathbf{k}_{\parallel}}$ the appropriate weights.

Fig.~\ref{fig:landauer}(a) illustrates the transmission for two cases: for the pristine Al wire the transmission is an integer staircase where $T(E)$ merely counts the number of bands crossing that energy. The
steps consequently coincide with the band edges and their heights reflect the band degeneracies. Inserting the H atom breaks the translational symmetry and backscattering at the junction causes the plateaus to collapse into a smooth non-integer curve, and sharp Fano-type dips appear where the hydrogen resonance interferes with the continuum of the wire. The finite-bias current follows as
\begin{equation}
\begin{split}
  I(V) &= \frac{2e}{h}\int_{-\infty}^{\infty}
    T(E)\bigl[f(E-\mu_{L}) - f(E-\mu_{R})\bigr]\,dE, \\
  \mu_{L/R} &= E_{F} \pm eV/2,
\end{split}
\label{eq:transport:current}
\end{equation}
evaluated with the zero-bias transmission, that is, neglecting the self-consistent rearrangement of the potential under applied bias. Fig.~\ref{fig:landauer} (b) shows the current $I(V)$ and the differential conductance $dI/dV$ which maps the transmission features onto the bias window. The conductor Green's function additionally yields the spectral density of states projected on the conductor region,
\begin{equation}
  N(E) = -\frac{1}{\pi}\sum_{\mathbf{k}_{\parallel}}
         w_{\mathbf{k}_{\parallel}}\,
         \mathrm{Im}\,\mathrm{Tr}\,
         G^{r}(\mathbf{k}_{\parallel},E),
  \label{eq:transport:dos}
\end{equation}
which is useful in assigning transmission features to specific conductor states.

\section{Lattice dynamics and the phonopy workflow}
\label{sec:phonons}

Vibrational properties are computed within \paoflow{} through a tight
integration with the \textsc{phonopy} package~\cite{Togo2015,Togo2023}, exposed
by the \code{phonon\_setup} and \code{phonons} drivers.  The design mirrors the
rest of the ecosystem: the \emph{same} unit cell that defines the PAO
Hamiltonian is bridged to a \code{phonopy.Phonopy} object
(\code{calculator=\textquotesingle qe\textquotesingle}), so the electronic and
the vibrational descriptions of a material are generated from a single,
consistent structural input.

\subsection{Finite-displacement force constants}
\label{sec:phonons:fc}

Harmonic phonons are obtained by the finite-displacement (frozen-phonon)
method.  Given a user-specified supercell matrix, \code{phonopy} enumerates the
symmetry-inequivalent atomic displacements; \paoflow{} then writes a complete,
ready-to-run \code{pw.x} SCF input for each displaced supercell
(\code{supercell-NNN.in}).  After the DFT runs, the Hellmann-Feynman forces are
harvested and the second-order (interatomic) force constants are assembled from
finite differences,
\begin{equation}
  \Phi_{i\alpha,\,j\beta} \;=\;
    \frac{\partial^2 E}{\partial u_{i\alpha}\,\partial u_{j\beta}}
  \;\simeq\; -\,\frac{\partial F_{j\beta}}{\partial u_{i\alpha}},
  \label{eq:phonon:fc}
\end{equation}
where $u_{i\alpha}$ is the displacement of atom $i$ along Cartesian direction
$\alpha$ and $F_{j\beta}$ the force on atom $j$.  The dynamical matrix and the
phonon dispersion follow from the mass-weighted lattice Fourier transform,
\begin{multline}
  D_{i\alpha,\,j\beta}(\mathbf{q}) =
  \frac{1}{\sqrt{M_i M_j}}
  \sum_{\vR} \Phi_{i\alpha,\,j\beta}(\vR)\,
  e^{\,i\mathbf{q}\cdot\vR},
 \\
  D(\mathbf{q})\,\mathbf{e}_{\mathbf{q}\nu}
  = \omega_{\mathbf{q}\nu}^{2}\,\mathbf{e}_{\mathbf{q}\nu},
  \label{eq:phonon:dynmat}
\end{multline}
yielding the frequencies $\omega_{\mathbf{q}\nu}$, the density of states, and
the harmonic thermodynamic functions (free energy, entropy, heat capacity).
The workflow operates in two phases -- \emph{generate} (write the displaced
inputs) and \emph{analyse} (ingest the forces) -- and is fully compatible with
the DFT+U(+V) electronic structure produced by ACBN0: an on-site
\code{HUBBARD} card is propagated to every displaced supercell so that the
forces, and hence the force constants, are consistent with the corrected
ground state.

\subsection{Polar corrections and derived spectra}
\label{sec:phonons:nac}

For polar insulators the non-analytical correction (NAC) restores the
LO-TO splitting near $\Gamma$ through the long-range dipole term
\begin{equation}
  D^{\mathrm{NA}}_{i\alpha,\,j\beta}(\mathbf{q}\!\to\!0) \propto
  \frac{\big(\mathbf{q}\cdot\mathbf{Z}^{*}_{i}\big)_{\alpha}\,
        \big(\mathbf{q}\cdot\mathbf{Z}^{*}_{j}\big)_{\beta}}
       {\mathbf{q}\cdot\boldsymbol{\varepsilon}^{\infty}\cdot\mathbf{q}},
  \label{eq:phonon:nac}
\end{equation}
built from Born effective charges $\mathbf{Z}^{*}_{i}$ and the
high-frequency dielectric tensor $\boldsymbol{\varepsilon}^{\infty}$, either
imported from a density functional perturbation theory calculation or computed from a central difference of forces and polarization under an electric field perturbation.  From the
zone-centre eigenvectors and $\mathbf{Z}^{*}$ the code evaluates the mode
effective charges and infrared oscillator strengths, the (non-resonant) Raman
activities, and the ionic contribution to the dielectric function,
\begin{equation}
  \varepsilon_{\alpha\beta}(\omega) = \varepsilon^{\infty}_{\alpha\beta}
  + \sum_{\nu}
    \frac{S_{\nu,\alpha\beta}}
         {\omega_{\nu}^{2}-\omega^{2}-i\,\omega\,\gamma_{\nu}},
  \label{eq:phonon:eps}
\end{equation}
with $S_{\nu}$ the mode oscillator-strength tensor.  Anharmonic thermal
expansion and the temperature dependence of the elastic and thermodynamic
response are accessible through the quasi-harmonic approximation (QHA), in which
the Helmholtz free energy is expressed as $F(V,T)=E(V)+F_{\mathrm{vib}}(V,T)$.

\section{Electron-phonon coupling from pseudo-atomic-orbital interpolation}
\label{sec:elphon}

Accurate first-principles electron-phonon coupling (EPC) demands dense Brillouin-zone sampling to resolve the strongly momentum-dependent scattering phase space in the vicinity of the Fermi surface. As a result, quantities such as the pairing interaction strength $\lambda$ can be acutely sensitive to the underlying $\bf{k}$- and $\bf{q}$-mesh densities. State-of-the-art workflows, including EPW \cite{Giustino2017} and Perturbo \cite{Perturbo2021}, address this bottleneck through Wannier interpolation, enabling efficient evaluation of electron-phonon matrix elements on ultra-dense grids. Here we instead implement, following Agapito and Bernardi~\cite{AgapitoBernardi2018}, an interpolation of the density-functional perturbation theory (DFPT) electron-phonon vertex in the pseudo-atomic-orbital (PAO) gauge to compute the isotropic Eliashberg function $\alpha^{2}F(\omega)$ and derived EPC metrics. Unlike Wannier-based schemes, the PAO gauge is fixed and smooth by construction, eliminating the need for band disentanglement or gauge fixing.

Starting from a non-self-consistent calculation on the full (unreduced)
Monkhorst-Pack grid and the DFPT coupling $d_{mn,\kappa\alpha}(\bf{k},\bf{q})$
on the same coarse mesh, we rotate each matrix element into the PAO basis,
\begin{equation}
  g^{\mathrm{PAO}}_{ij,\kappa\alpha}(\bf{k},\bf{q}) =
  \bigl[A^{\dagger}_{\bf{k}+\bf{q}}\, d_{\kappa\alpha}(\bf{k})\, A_{\bf{k}}\bigr]_{ij},
\end{equation}
where $A_{\bf{k}}$ are the \paoflow{} projection operators as defined in Sec. \ref{sec:pao_ham}. Fourier
transforming the electron momentum $\bf{k}\!\to\!\bf{R}_e$ and the phonon
momentum $\bf{q}\!\to\!\bf{R}_p$ yields the double real-space vertex
$g_{ij,\kappa\alpha}(\bf{R}_e,\bf{R}_p)$, from which the coupling at an arbitrary
pair $(\bf{k},\bf{q})$ is recovered by Wigner-Seitz back-transformation. The PAO
Hamiltonian $H(\bf{k})$ is interpolated onto a dense $\bf{k}$ grid by the same
generalized Fourier procedure, while the dynamical matrices are interpolated to
a dense $\bf{q}$ grid with the enforced acoustic sum rule, giving the mode
frequencies $\omega_{\bf{q}\nu}$ and polarization vectors.

The mode-resolved coupling is obtained from the Fermi-surface double delta,
\begin{equation}
  \lambda_{\bf{q}\nu} =
  \frac{1}{N(E_F)\,\omega_{\bf{q}\nu}^{2}}\,
  \frac{1}{N_{k}}\sum_{\bf{k}}
  \bigl|g_{\bf{q}\nu}(\bf{k})\bigr|^{2}\,
  \delta(\varepsilon_{\bf{k}})\,\delta(\varepsilon_{\bf{k}+\bf{q}}),
  \label{eq:lamqv}
\end{equation}
where energies are measured relative to $E_F$, so both electronic states lie on the Fermi surface. Brillouin-zone averaging of $\lambda_{\bf{q}\nu}$ over phonon branches and irreducible $\bf{q}$ points, weighted by the corresponding star weights $w_{\bf{q}}$, yields the Eliashberg spectral function $\alpha^{2}F(\omega)$, the total electron-phonon coupling $\lambda$, and the logarithmic average frequency $\langle\omega_{\log}\rangle$ that sets the characteristic phonon scale for pairing:
\begin{equation}
  \begin{aligned}
    \alpha^{2}F(\omega) &= \tfrac{1}{2}\sum_{\bf{q}\nu} w_{\bf{q}}\,
      \lambda_{\bf{q}\nu}\,\omega_{\bf{q}\nu}\,\delta\bigl(\omega-\omega_{\bf{q}\nu}\bigr),\\
    \lambda &= 2\!\int_{0}^{\infty} \frac{\alpha^{2}F(\omega)}{\omega}\,d\omega,\\
    \langle\omega_{\log}\rangle &= \exp\left( \frac{2}{\lambda} \int_{0}^{\infty} \frac{\alpha^{2}F(\omega)\,\ln\omega}{\omega}\,d\omega \right)
  \end{aligned}
\end{equation}

Because $\lambda_{\bf{q}\nu}$ is invariant under the crystal point group, the
dense $\bf{q}$ integration is restricted to the irreducible wedge (with time
reversal), reducing the number of evaluated points by up to the order of the
point group. The per-$\bf{q}$ work is distributed over MPI ranks, and the large real-space vertex
is held in a single node-shared copy. 

Table~\ref{tab:elphon-convergence} benchmarks the convergence of $\lambda$ and $\langle\omega_{\log}\rangle$ for non-relativistic fcc Pb (see Sec.~\ref{sec:computation} for computational details); reproducing prior DFPT results obtained without SOC \cite{Chulkov2012} requires fully converged values from at least $36^{3}$ points in both the $\bf{k}$ and $\bf{q}$ meshes. Notably, the PAO interpolation greatly simplifies the EPC workflow, makes these ultra-dense-mesh calculations computationally very inexpensive, and enables the evaluation of a plethora of properties derived from the knowledge of the EPC \cite{MBNEPCinprep}.

\begin{table}[htbp]
\centering
\begin{tabular*}{\linewidth}{@{\extracolsep{\fill}} rr S[table-format=1.4] S[table-format=2.1] }
\toprule
$N_k$ & $N_q$ & {$\lambda$} & {$\langle\omega_{\log}\rangle$ (K)} \\
\midrule
12 & 6  & 2.0404 & 61.3   \\
18 & 6  & 1.4857 & 60.3   \\
18 & 18 & 1.2562 & 61.3   \\
24 & 12 & 1.1779 & 62.0   \\
24 & 24 & 1.1384 & 60.5   \\
36 & 18 & 1.0758 & 61.2   \\
36 & 36 & 1.0829 & 59.4   \\
42 & 42 & 1.0832 & 59.0   \\
\addlinespace
\bottomrule
\end{tabular*}
\caption{Convergence of the electron-phonon coupling constant $\lambda$ and the logarithmic average frequency for FCC Pb starting from a DFPT calculation with a $9^3$ {\bf k}-grid and $6^3$ {\bf q}-grid.}
\label{tab:elphon-convergence}
\end{table}
\section{Electronic structure and topology}
\label{sec:topological}
\subsection{Fermi surfaces and quantum oscillations}
\label{sec:topo:skeaf}

\paoflow{} provides tools for the analysis of Fermi surfaces and quantum oscillatory phenomena. In the presence of a strong magnetic field, electronic states are quantized into Landau levels, corresponding to closed orbits defined by the intersections of the Fermi surface with planes perpendicular to the magnetic field direction. The Landau quantization gives rise to quantum oscillatory phenomena, such as de Haas-van Alphen and Shubnikov-de Haas effects, whose oscillation frequencies are determined by the extremal cross-sectional areas of the Fermi surface. Determining these extremal areas provides a direct connection between electronic structure calculations and experimentally measurable quantities~\cite{Shoenberg1984, LifshitzKosevich1956, Onsager1952_dHvA}.

To enable automated quantum-oscillation calculations, the \textsc{PySKEAF} module of \paoflow{} implements the widely adopted SKEAF (Supercell $\mathbf{k}$-space Extremal Area Finder) algorithm of Rourke and Julian~\cite{Rourke2012}. The PAO Hamiltonians interpolated on dense $\mathbf{k}$-meshes provide high-resolution Fermi surfaces that are analyzed directly by \textsc{PySKEAF}. This integrated workflow identifies extremal Fermi-surface orbits and computes the associated oscillation frequencies and cyclotron effective masses without additional post-processing.

For a selected Fermi energy and direction of magnetic field $\bf{B}$, the \textsc{PySKEAF} algorithm:

\begin{enumerate}
    \item constructs planes perpendicular to the magnetic field, determines their intersections with the Fermi surface to generate two-dimensional Fermi contours, and tracks corresponding contours across adjacent planes to identify extremal closed orbits;
    \item computes the quantum oscillation frequency $F$ in $1/B$ space from the Onsager relation~\cite{Onsager1952_dHvA}:
    \begin{equation}
        F=
        \frac{\hbar}{2\pi e}
        A_{\mathrm{ext}}
        \label{eq:onsager}
    \end{equation}
    where $A_{\mathrm{ext}}$ is the extremal cross-sectional area of the Fermi surface;
    \item calculates the cyclotron effective mass according to
    \begin{equation}
        m_c^{*}
        =
        \frac{\hbar^{2}}{2\pi m_e}
        \left|
        \frac{\partial A_{\mathrm{ext}}}{\partial E}
        \right|_{E=E_F}
        \label{eq:cyclotron_mass}
    \end{equation}
    and determines the electron- or hole-like character of each extremal orbit from the geometry of the Fermi surface contour.
\end{enumerate}

\begin{figure}[!h]
    \centering
    \includegraphics[width=\linewidth]{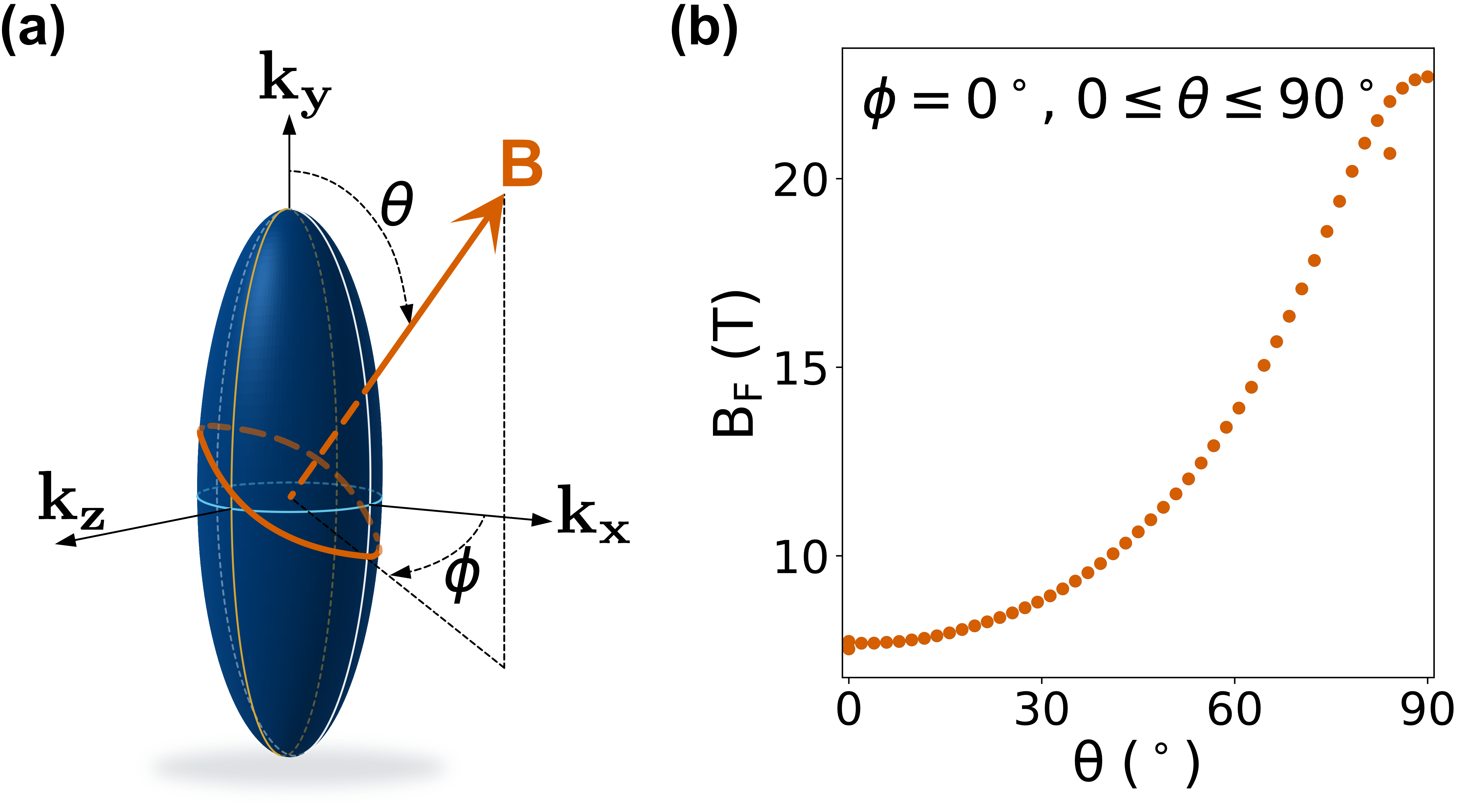}
    \caption{Example of Fermi surface and Shubnikov-de Haas frequencies in bulk ZrTe$_5$ calculated using \paoflow{}. Original data taken from Ref.~\cite{Chen2025}, specifically the case of weak topological insulator at hole doping level $\mathrm{E-E_{VBM} = -11.5 \; meV}$. (a)~Calculated Fermi surface with an extremal cross-sectional area perpendicular to the magnetic field $\vb{B}$ (orange). (b)~The corresponding Shubnikov-de Haas oscillation frequencies when rotating the magnetic field $\vb{B}$ in the $\mathrm{k_x-k_y}$ plane.}
    \label{fig:qo}
\end{figure}

The analysis is repeated automatically for different polar and azimuthal angles of the magnetic field, providing the angular dependence of the quantum oscillation frequencies. An example of using the workflow is shown in Fig.~\ref{fig:qo}, where the Fermi surface of ZrTe$_5$ and the quantum oscillations are calculated in \paoflow{}~\cite{Chen2025}.
\subsection{$\mathbb{Z}_2$ topological invariants}
\label{sec:topo:z2}


The $\mathbb{Z}_{2}$ number
 of a time-reversal-invariant insulator can be
computed in two complementary ways, both acting on the same PAO
Hamiltonian. The first is the Pfaffian construction of Fu, Kane, and
Mele~\cite{FuKane2007,FuKaneMele2007}. In this approach, one constructs
the sewing matrix
$w_{mn}(\mathbf{k}) = {u_m(-\mathbf{k})}{\Theta}{u_n(\mathbf{k})}$,
with $\Theta$ denoting the antiunitary time-reversal operator.
At each time-reversal-invariant momentum (TRIM) $\Gamma_i$, four in two
dimensions and eight in three dimensions, this matrix is antisymmetric.
When restricted to the occupied manifold, it yields the sign entering the
$\mathbb{Z}_{2}$ invariant,
\begin{equation}
  \delta_i \;=\;
  \frac{\mathrm{Pf}\bigl[w(\Gamma_i)\bigr]}
       {\sqrt{\det w(\Gamma_i)}} \;=\; \pm 1,
  \qquad
  (-1)^{\nu_0} \;=\; \prod_{i} \delta_i ,
  \label{eq:pfaffian}
\end{equation}
where the product runs over all TRIMs for the strong index $\nu_0$ and
over the four TRIMs of each $k_j = \pi$ plane for the weak indices
$\nu_1,\nu_2,\nu_3$. This procedure is computationally efficient because
it requires only the wave functions at the TRIMs. However, it must be
applied with care, since the result depends on a consistent gauge choice
for the occupied subspace.

In the second approach, the PAO Hamiltonian is exported to
\textsc{Z2Pack}~\cite{Gresch2017,Soluyanov2011}, which tracks the evolution of
the hybrid PAO charge centers, equivalent to the Wannier charge centers,
along a family of loops spanning half of the Brillouin zone. The
invariant is then determined from the parity of the number of times the
largest gap between the charge centers is crossed. This criterion is
explicitly gauge invariant and requires no symmetry beyond time reversal,
at the cost of many diagonalizations over adaptively refined loops.

\subsection{Chern numbers and Chern insulators}
\label{sec:topo:chern}
The Chern number $C$ is an integer-valued topological invariant for systems that break time-reversal symmetry. It is defined as the
Brillouin-zone integral of the Berry curvature summed over the occupied Bloch bands, which encodes the global geometry of the corresponding band
manifold~\cite{thouless1982,haldane1988,xiao2010}
\begin{equation}
\label{eq:chern1}
    C = \frac{1}{2\pi} \sum_{n \in \mathrm{occ}} \int_{\mathrm{BZ}}
    \Omega_{n}(\mathbf{k}) \, d^{2}k . 
 \end{equation}

A nonzero $C$, which requires broken time-reversal symmetry, identifies a Chern-insulating phase in which the anomalous Hall conductivity is quantized as $\sigma_{xy} = C e^{2}/h$ and $|C|$ chiral edge modes traverse the bulk gap. 

A vanishing Chern number, however, does not imply topological triviality: time-reversal-invariant systems have $C = 0$ by construction, yet may host nontrivial phases protected by time-reversal or crystalline symmetries, whose characterization requires other invariants.

\subsection{Mirror Chern numbers}
\label{sec:topo:mirror_chern}

Among additional invariants, the mirror Chern number plays a central role in crystals possessing a mirror reflection symmetry~\cite{teo2008,fu2011}. On the planes in the reciprocal space that are invariant to mirror operation $\mathcal{M}$, the Hamiltonian satisfies $[\mathcal{H}(\kk),\mathcal{M}]=0$, and since $\mathcal{M}^{2}=-1$ for spin-$\tfrac12$ electrons, the Bloch states may be labelled by the mirror eigenvalues $\pm i$. The occupied manifold thus separates into two independent sectors, each carrying an ordinary Chern number $C_{\pm i}$, from which the total and mirror Chern numbers are defined as $C=C_{+i}+C_{-i}$ and $C_{M}=(C_{+i}-C_{-i})/2$, respectively. In the presence of time-reversal symmetry
$C_{+i}=-C_{-i}$, forcing $C$ to vanish while allowing $C_{M}$ to remain finite; the mirror Chern number therefore captures topological information inaccessible to the ordinary Chern invariant. A nonzero $C_{M}$ identifies a mirror-protected topological crystalline phase and gives rise to symmetry-protected gapless boundary states on surfaces or edges that preserve the corresponding mirror symmetry~\cite{fu2011,hsieh2012}.

For a two-dimensional crystal the horizontal mirror $\Mz:z\to-z$ is of particular relevance, since its invariant plane coincides with the entire Brillouin zone. When $\Mz$ commutes with $\mathcal{H}(\kk)$, the sector decomposition applies throughout the zone, providing a direct route to the evaluation of $C_{M}$~\cite{teo2008}. In the present calculations $C_{M}$ is obtained from the PAO Hamiltonian, with the sector-resolved Chern numbers computed using \textsc{Z2Pack}~\cite{Gresch2017}.



\subsection{Weyl points}
\label{sec:topo:weyl}
\begin{figure*}[!ht]
    \centering
    \includegraphics[width=\textwidth]{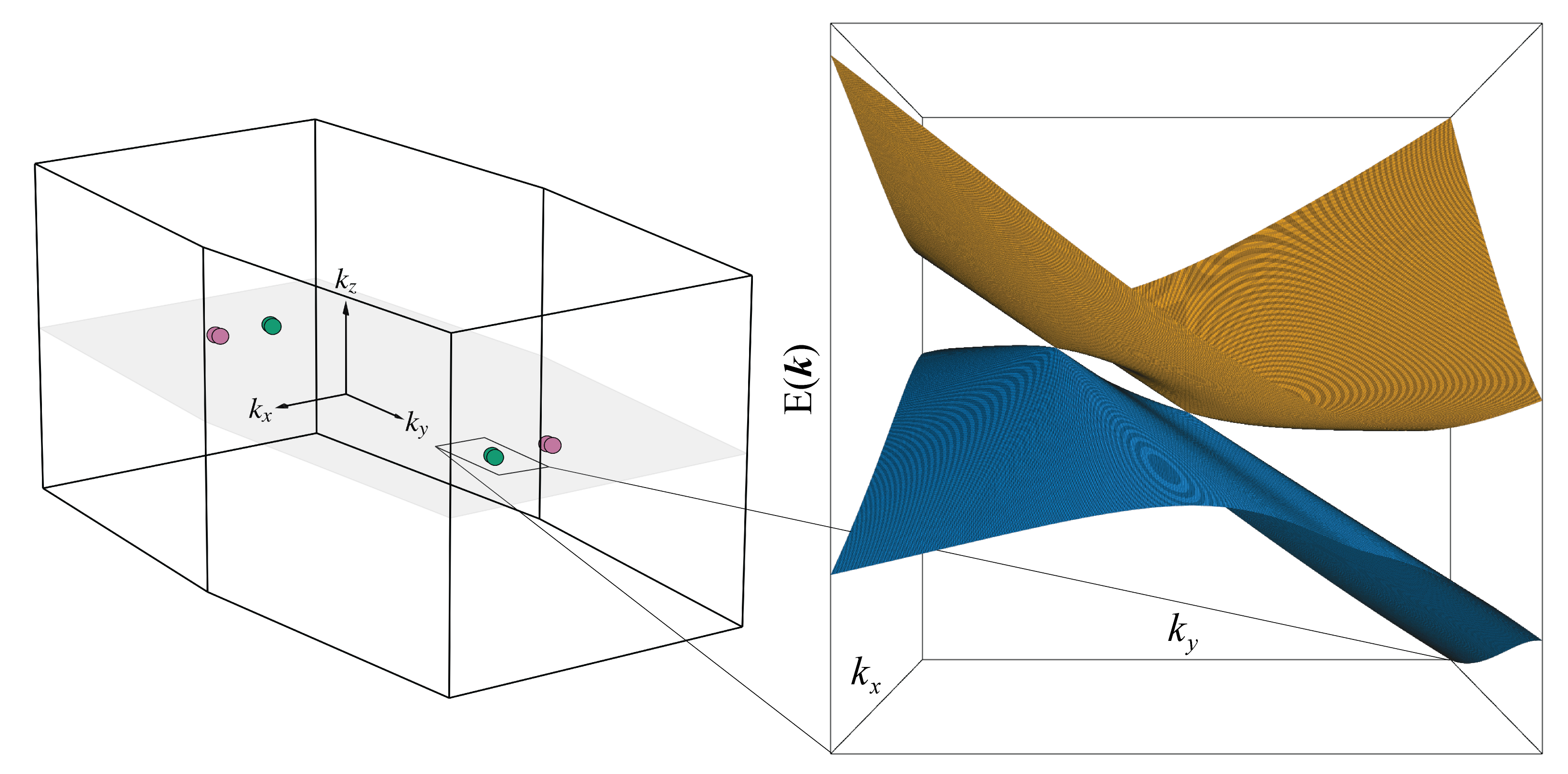}
    \caption{Example of Weyl point identification in MoP$_2$ using \paoflow{}. Left: Calculated Weyl points in the Brillouin zone. The green  and purple colors correspond to opposite chiralities. Right: Electronic band structure $E (k_x,k_y)$ around two neighboring Weyl points, demonstrating the linear crossing of the bands (The the $k_x,k_y$ plane spans an area of 0.00562 1/\AA$^2$ and an energy range from -0.622 eV to -0.196 eV).}
    \label{fig:WPs}
\end{figure*}
Weyl points are topologically protected linear crossings of two bands in three-dimensional reciprocal space, characterized by an integer topological charge $\chi=\pm1$~\cite{Wan2011,Vanderbilt2018}. \paoflow{} identifies the locations of Weyl points by searching for vanishing band gaps throughout the Brillouin zone. The Brillouin zone is first divided into a coarse grid of rectangular boxes, within which the energy gap between the two bands of interest is locally minimized using the L-BFGS-B optimization algorithm implemented in SciPy. The chirality of the resulting candidate band crossings is subsequently verified using the external Python library Z2Pack~\cite{Gresch2017}, which determines the Chern number of a spherical surface enclosing each candidate point. As an additional consistency check, the total chiral charge in the BZ must vanish, ensuring that Weyl points always occur in pairs of opposite chirality. An example of the calculation is shown in Fig.~\ref{fig:WPs}, illustrating the Weyl points identified in MoP$_2$.

\subsection{Band unfolding}
\label{sec:topo:unfold}
\begin{figure*}[!ht]
  \centering
  \includegraphics[width=\textwidth]{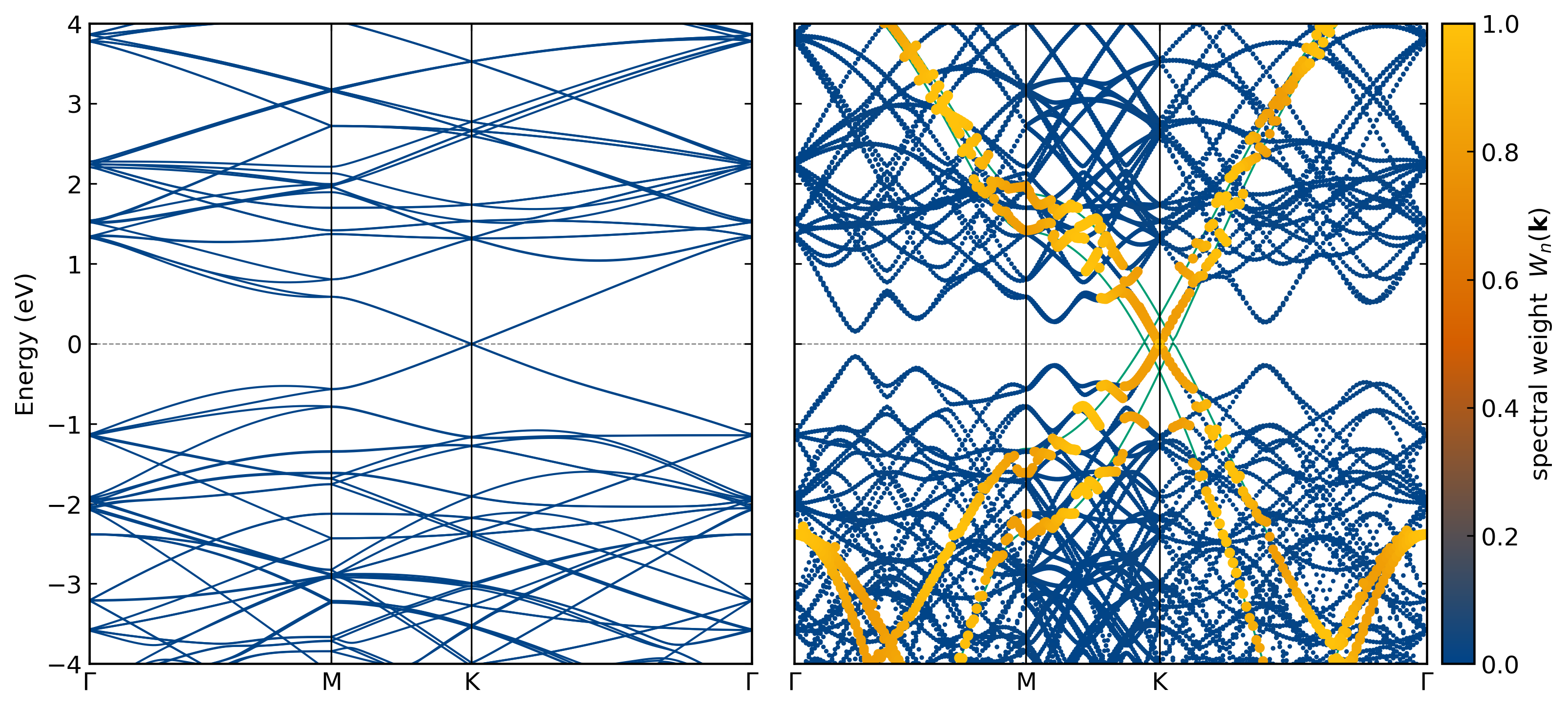}
  \caption{Folded (SC, left) and unfolded (PC, right) band structures for twisted bilayer graphene in the $(3,2)$ moir\'{e} supercell. In the unfolded plot, the color intensity is proportional to the spectral weight $w_n(\mathbf{k})$ [Eq.~\eqref{eq:spectral_weight}], highlighting the contribution of SC eigenstates to effective PC bands.}
  \label{fig:unfolded}
\end{figure*}
Band unfolding restores a primitive-cell (PC) representation of electronic bands from a supercell (SC) calculation. This is particularly useful for defects, surfaces, heterostructures, and moir\'{e} superlattices, where the enlarged real-space periodicity folds the SC bands into a reduced Brillouin zone and obscures direct comparison with PC dispersions.

\paoflow{} performs unfolding using the spectral-weight formalism~\cite{Boykin_2005,Popescu2012,Nishi_2017}. Given an integer, invertible transformation matrix $M$ relating the lattice vectors,
$\mathbf{A}_{\text{SC}} = M\,\mathbf{A}_{\text{PC}}$,
its determinant gives the number of PC translations in the SC, $N = |\det M|$. Denoting by $\{\mathbf{R}_\ell\}_{\ell=1}^{N}$ the corresponding translation vectors, the spectral weight associated with PC wavevector $\mathbf{k}$ for SC eigenstate $(n,\mathbf{K})$ is
\begin{equation}
  w_n(\mathbf{k}) = \frac{1}{N} \sum_{\alpha}\sum_{m}
    \left| \sum_{\ell=1}^{N} e^{-2\pi i \mathbf{k}\cdot\mathbf{R}_\ell}\,
    C^{n}_{\alpha, m, \ell}(\mathbf{K}) \right|^2,
  \label{eq:spectral_weight}
\end{equation}
where $C^{n}_{\alpha,m,\ell}(\mathbf{K})$ is the SC eigenvector coefficient on orbital $m$ of the atom obtained by translating the PC atom $\alpha$ by $\mathbf{R}_\ell$. The SC wavevector $\mathbf{K}$ maps onto the PC wavevector $\mathbf{k}$ through folding,
$\mathbf{K} = \mathbf{k} M^{-1} \pmod{\mathbf{G}_{\text{PC}}}$.

The unfolded spectral function follows as
\begin{equation}
  A(\mathbf{k},E) = \sum_n w_n(\mathbf{k})\,\delta(E-\varepsilon_{n\mathbf{K}}),
  \label{eq:spectral_function}
\end{equation}
so that $w_n(\mathbf{k})$ directly controls the intensity of the unfolded bands along a chosen PC path.

Figure~\ref{fig:unfolded} illustrates the procedure for twisted bilayer graphene in the $(3,2)$ moir\'{e} cell (76 atoms), comparing the folded SC bands with the unfolded PC representation. The unfolded weights emphasize the PC-like components of the SC eigenstates and reproduce the low-energy features reported in Ref.~\cite{Nishi_2017}. Here we use the same EDTB model of Fig. \ref{fig:TBG} (see also Ref. \cite{Marco2026EDTB}).

\section{Computational details}
\label{sec:computation}

 Silicon and platinum calculations (Fig. \ref{fig:basis} and \ref{fig:acbn0} (left)) have been carried out using Quantum ESPRESSO, with input files generated using \texttt{paoflow-gen-qe}, a command line generator that builds a Quantum ESPRESSO \texttt{scf} input from an AFLOW database entry (\url{https://aflow.org})~\cite{2012-Curtarolo-CMS-58-227} and is shipped with the \paoflow{} distribution. The generated input file inherits the geometry and k-meshes from the database entries (Si AUID: aflow:08ab41c5f54850db, Pt AUID: aflow:0498f57236cee872) while pseudopotentials are chosen from the Pseudo Dojo repository (\url{https://www.pseudo-dojo.org/})~\cite{pseudodojo2018}. For convenience, \paoflow{} ships a \texttt{PSEUDOS/} folder that contains Perdew-Burke-Ernzerhof (PBE) scalar-relativistic and fully-relativistic pseudopotentials with the corresponding optimal energy cutoffs to automate the input file generation. Input for the MoS$_2$ system (Fig. \ref{fig:acbn0}) has been generated using \texttt{paoflow-gen-qe} on the corresponding entry in the C2DB database (\url{https://c2db.fysik.dtu.dk/material/1MoS2-1})~\cite{Haastrup_2018,Gjerding_2021} with a $18\times18\times1$ {\bf k}-grid and scalar relativistic pseudopotentials from Pseudo Dojo. 

The DFT calculations shown in Fig. \ref{fig:boltzmann} were performed for bulk silicon, using a fcc unit cell with $a=5.609 $ \AA. Calculations were carried out using Quantum ESPRESSO with the PBE functional~\cite{pbe_gga}, using scalar relativistic projector-augmented wave pseudopotential from the \textsc{PSLIBRARY} database. Plane wave cut-off energy was set to 70 Ry and the energy convergence threshold to $10^{-9}$ Ry. A Monkhorst-Pack $k$-mesh of $16\times16\times16$ and $32\times32\times32$ was used for self-consistent and non-self-consistent steps, respectively~\cite{Monkhorst1976}.

Calculations for NiTa$_3$S$_6$ (Fig. \ref{fig:kuboshe}) were carried out using the Vienna Ab initio Simulation Package (VASP)~\cite{Kresse1996,Kresse1999}. Exchange and correlation were described using the PBE functional. A plane-wave energy cutoff of 350 eV was used, with the total energy converged to $10^{-7}$ eV. BZ integrations were performed on a $16\times16\times8$ Monkhorst-Pack $k$-point mesh with a Gaussian smearing of 0.05 eV. SOC was included self-consistently in the DFT calculations. We used a hexagonal unit cell, containing two Ni atoms, six Ta atoms, and twelve S atoms, with lattice constants $a=b=5.77$~\AA\ and $c=12.03$~\AA. The ionic positions were relaxed until the residual forces were below $10^{-3}$ eV/\AA. In \paoflow{}, we used the basis set: Ni [4s, 4p, 3d]; Ta [6s, 5p, 6p, 5d]; Nb [5s, 3p, 4p, 5p, 3d, 4d, 4f]; and S [3s, 3p, 3d] and $k$-mesh of $60\times60\times40$ for the interpolation of the Hamiltonian. Further details of the calculation are provided in Ref.~\cite{Tenzin2025}.

The DFT calculations for the NbIrTe$_4$ tetralayer system were performed with Quantum ESPRESSO. The PBE functional, along with fully relativistic PAW pseudopotentials~\cite{Kresse1999}, was employed. A plane-wave energy cutoff of 80 Ry was used, and the energy convergence threshold was set to $10^{-9}$ Ry. The lattice parameters of the orthorhombic cell were set to $a=5.77$~\AA\ and $b=12.03$~\AA\ based on experimental findings~\cite{Lee2024}, along with a vacuum layer of 60 \AA. We used a Monkhorst-Pack mesh of $16\times 10\times1$ in DFT, and in order to ensure convergence, the $k$-mesh was increased to $80\times 40\times1$ via interpolation for the spin Hall conductivity calculations.

The DFT calculations for quantum transport were performed on a monatomic Al wire hosting a single H impurity, using Quantum ESPRESSO with the PBE functional. Al was described by a norm-conserving RRKJ pseudopotential and H by an ultrasoft Vanderbilt pseudopotential, with a plane-wave cutoff of 18 Ry for the wavefunctions and a self-consistency threshold of $10^{-8}$ Ry. BZ integrations used a $1\times1\times4$ Monkhorst-Pack mesh~\cite{Monkhorst1976} shifted along the transport direction. Transport quantities were evaluated within \paoflow{} on an energy grid of 9001 points spanning $[-7,2]$~eV with respect to the Fermi level, using a broadening of $\eta=0.5$~meV. The pristine wire was treated as a periodic (bulk) conductor, while the H-decorated junction was partitioned into a left-conductor-right geometry with three principal layers on each side acting as leads along $z$. The current was obtained by integrating the transmission over the bias window for $V\in[-1,1]$~V sampled on 1500 points.

The Fermi surface and SdH oscillation frequencies of ZrTe$_5$ shown in Fig. \ref{fig:qo} were derived from DFT calculations using Quantum ESPRESSO with the PBE functional. The ion-electron interactions were treated using fully relativistic projector-augmented wave pseudopotentials taken from the \textsc{PSLIBRARY} database~\cite{pp_2014}. The plane-wave cutoff energy was set to $80 \,$ Ry, and the Monkhorst-Pack mesh was $16 \times 16 \times 8$ for the self-consistent  and $30 \times 30 \times 10$ for the non-self-consistent calculations. We used a primitive cell of the base-centered orthorhombic crystal; the lattice vectors are $\vb{a} = \qty(7.73422, 0, 0)$, $\vb{b}=\qty(-6.6711, 3.91337, 0)$ and $\vb{c} = \qty(0, 0, 13.709)$ (\AA\ units). SOC was taken into account in all the calculations, except during atomic structural optimizations. In \paoflow{}, the Hamiltonian was interpolated on denser $k$ grids of $100 \times 100 \times 18$ to generate an accurate and smooth Fermi surface. The calculation of SdH oscillation frequencies in the \textsc{PySKEAF} module of \paoflow{} was performed using a Fermi energy tag of $-0.2$ eV to achieve the doping energy level $\mathrm{E-E_{VBM} = -11.5}$ meV, and 350 interpolated points per single side were considered. \textsc{PySKEAF} uses the conventional orthogonal $\vb{k}$ coordinate for the direction of $\vb{B}$-field, but the coordinate of a Fermi surface is written in reciprocal basis inside a BXSF file. Therefore, one must be careful about setting the angles of the $\vb{B}$-field. The result illustrated in Fig.~\ref{fig:qo}(b) was achieved by fixing the polar angle to $90^{\circ}$ and rotating the azimuthal angle from $164.802^{\circ}$ til $254.802^{\circ}$ in \textsc{PySKEAF}, followed by a transformed reference frame in Fig.~\ref{fig:qo}(a) for better illustration.

The DFT calculations for MoP$_2$ were performed using Quantum ESPRESSO with the PBE functional. Fully relativistic norm-conserving pseudopotentials were employed. The plane wave cut-off energy was set to 50 Ry, and the energy convergence threshold to $10^{-8}$ Ry. A Monkhorst-pack $k$-mesh of $18 \times 18 \times 16$ was used. An orthorhombic base-centered cell was used, with the lattice vectors set to $\vb{a} = \qty(1.572100, 5.590044, 0)$, $\vb{b}=\qty(-1.572100, 5.590044, 0)$ and $\vb{c} = \qty(0, 0, 4.981609)$ (\AA\ units).
For the calculation of the band structure, in order to differentiate the two neighboring Weyl points, a small area of 0.00562 1/\AA $^2$, centered on the Weyl point, was densely interpolated on a grid of $500 \times 500$ $k$-points.   

For the EPC workflow in Pb we ran Quantum ESPRESSO with a norm-conserving PBE pseudopotential from the PSLIBRARY with a kinetic energy cutoff of 30~Ry. A $9^{3}\,\bf{k}$-grid was used for the initial DFT calculation and a $6^{3}\,\bf{q}$-grid for the DFPT calculation prior to interpolation. While SOC is essential for a quantitatively faithful description of Pb\cite{Chulkov2012}, we intentionally adopt the nonrelativistic limit here to enable a stringent benchmark against established DFPT reference data.

\section{Software architecture}
\label{sec:architecture}

\paoflow{} is organized as a layered, performance-portable software stack: an MPI-parallel Python layer (NumPy/SciPy) provides the high-level API, data flow, and workflow orchestration, while performance-critical kernels can be transparently offloaded to an optional Rust backend. This separation of concerns keeps the user-facing interface stable, improves testability and maintainability, and enables targeted optimization without constraining downstream workflows.

To address memory and scaling bottlenecks, \paoflow{} includes a dedicated sparse-algebra execution path designed for large supercells or moir\'e systems constructed with the environment-dependent tight-binding scheme. In the standard PAO workflow dense matrices scale quadratically with the number of atomic orbitals and quickly become the dominant resource constraint. The sparse backend mitigates this by storing the real-space Hamiltonian as a thresholded list of interacting orbital pairs (with approximately linear growth in storage) and by avoiding construction of the full $\mathbf{k}$-space Hamiltonian across the entire mesh: instead, it assembles and diagonalizes each $\mathbf{k}$ point on demand. Currently, band structures, densities of states, and Boltzmann transport coefficients are supported in this mode, with additional properties being integrated through the same backend-agnostic interfaces. 

Further details on the implementation, optimization, validation and verification are provided in the documentation section of the \paoflow{} website (\url{https://paoflow.org}) and in the Wiki of our GitHub repository (\url{https://github.com/marcobn/PAOFLOW/wiki}).

Portions of the code development, debugging, and generation of the software documentation were carried out with the assistance of GitHub Copilot (Claude Opus
4.6-8, Anthropic~\cite{Anthropic2025Claude4}), an AI-powered programming assistant integrated into the Visual Studio Code development environment. In these cases, all scientific decisions, including model design,
choice of physical approximations, parameter selection, and
interpretation of results, were made by the authors.

\section{Conclusions}
\label{sec:conclusions}
The developments presented in this work reflect the continuous evolution of \paoflow{} from a Hamiltonian interpolation package into a comprehensive ecosystem for first principles materials characterization. The recent development of the code has substantially expanded the accessible physical properties while maintaining a workflow based on compact PAO Hamiltonians. As demonstrated throughout this paper, the same PAO Hamiltonian can serve as the starting point for calculations of electronic structure, optical response, transport phenomena, and topological properties, allowing for direct comparison with an increasingly broad range of experimental techniques, and enabling the design of high-throughput workflows for scalable generation of training-quality data and AI-ready property datasets for materials discovery.

Beyond the methodological developments presented here, the modular architecture of \paoflow{} provides a flexible framework for adding new theoretical approaches as they emerge. Recent years have witnessed rapid progress in areas such as nonlinear transport, quantum geometry, spin-orbit phenomena, orbitronics, and topological materials, many of which rely on fundamental quantities already available in \paoflow{}. Extending the package towards these emerging directions therefore represents a natural continuation of its development. At the same time, extending \paoflow{} to additional DFT codes will remain an important direction for future development. In particular, extending support to all-electron and localized basis codes, such as WIEN2k and SIESTA, would broaden its applicability and facilitate its integration into diverse computational workflows.

\begin{acknowledgments}
The authors wish to  acknowledge the contributions of past developers who made this project possible: Franklin Cerasoli, who built the core architecture of the code and made it a modern modular software application; Luis Agapito, whose work laid the foundation for the projectability criterion and contributed to the development of the ACBN0 functional; Marco Fornari and Arrigo Calzolari, for support, encouragement and invaluable scientific discussions; Andrew Supka, for his deep understanding of computational structures and his contributions to the symmetry modules of \paoflow{}; and Haihang Wang, Laalitha Liyanage, Priya Gopal, Ilaria Siloi, Homayoun Jafari and Przemysław Przybysz for bearing with the barrage of testing new functions in the early stages of development and their scientific contributions stemming from the application of \paoflow{} to real life problems. 
M.B.N. has been supported, in part, by the US Department of Energy (DOE, Office of Basic Energy Sciences) under Grant No. DE-SC0024554, and the
National Science Foundation under Grant No. 2523217. J.S., S.A., M.F., Z.H., and K.T. acknowledge the funding from the European Research Council – ERC Consolidator Grant FERRERO. J.S. and C.C.Y. have been supported by the research program “Materials for the Quantum Age” (QuMat). This program (registration number 024.005.006) is part of the Gravitation program financed by the Dutch Ministry of Education, Culture and Science (OCW). J.S. acknowledges the Rosalind Franklin Fellowship from the University of Groningen. A.J. acknowledges funding from the NCCR MARVEL project funded by the Swiss National Science Foundation Grant No. 205602. M.C. acknowledges the financial support of CNPq Grant 317320/2021-1, FAPERJ Grant E26/200.240/2023, and INCT Materials Informatics.  S.C. acknowledges partial support from DoD-ONR (N00014-21-1-2515 and N00014-24-12768). The authors acknowledge the Texas Advanced Computing Center (TACC) at The University of Texas at Austin, the Extreme Science and Engineering Discovery Environment (XSEDE), the SURF Cooperative (The Netherlands) and Center for Information Technology of the University of Groningen for their support and for providing access to computational resources. Finally the authors acknowledge the support of the Santa Fe Institute for hosting the working group "Computational Frontiers in Quantum Materials" in May 2026, where many of the ideas and methods discussed in this paper took final shape.
\end{acknowledgments}

\bibliography{references}

\end{document}